\documentclass[conference,review, anonymous]{IEEEtran}
\let\labelindent\relax
\usepackage[linesnumbered,ruled,lined]{algorithm2e}
\usepackage{booktabs}
\usepackage{subcaption}
\usepackage[inkscapelatex=false]{svg}
\usepackage{multirow}
\usepackage{amsmath}
\usepackage{graphicx}
\usepackage{pifont}
\usepackage{makecell}
\usepackage{setspace}
\usepackage[numbers,sort&compress]{natbib}

\usepackage{enumitem}
\usepackage{amssymb}
\usepackage[normalem]{ulem}
\usepackage{tabularx}
\usepackage{tcolorbox}
\usepackage{threeparttable}
\usepackage{hyperref} 
\usepackage{cleveref} 

\hypersetup{colorlinks = true, 
	    linkcolor = blue, 
	    urlcolor = blue,
            citecolor = red} 
\ifCLASSINFOpdf

\else

\fi

\begin{document}
%
\title{\binenhance{}: An Enhancement Framework Based on External Environment Semantics for Binary Code Search}

\author{\IEEEauthorblockN{Yongpan Wang\IEEEauthorrefmark{2}\IEEEauthorrefmark{3},
Hong Li\IEEEauthorrefmark{2}\IEEEauthorrefmark{3}\IEEEauthorrefmark{1}, Xiaojie Zhu\IEEEauthorrefmark{4}, Siyuan Li\IEEEauthorrefmark{2}\IEEEauthorrefmark{3}, Chaopeng Dong\IEEEauthorrefmark{2}\IEEEauthorrefmark{3}, Shouguo Yang\IEEEauthorrefmark{5}, Kangyuan Qin\IEEEauthorrefmark{2}\IEEEauthorrefmark{3}
\thanks{\IEEEauthorrefmark{1}Corresponding Author}}
\IEEEauthorblockA{
\IEEEauthorrefmark{2}Institute of Information Engineering, Chinese Academy of Sciences, China\\ 
\IEEEauthorrefmark{3}School of Cyber Security, University of Chinese Academy of Sciences, China\\
\IEEEauthorrefmark{4}King Abdullah University of Science and Technology, Thuwal, Saudi Arabia\\
\IEEEauthorrefmark{5}Zhongguancun Laboratory, Beijing, China\\
ncepu\_wyp@163.com, \{lihong, lisiyuan, dongchaopeng, qinkangyuan\}@iie.ac.cn\\ xiaojie.zhu@kaust.edu.sa, yangshouguo@outlook.com\\}}

\newcommand{\binenhance}[0]{\textsc{BinEnhance}}
\newcommand{\blue}[1]{\textcolor{blue}{#1}}
  
\IEEEoverridecommandlockouts
\makeatletter\def\@IEEEpubidpullup{6.5\baselineskip}\makeatother
\IEEEpubid{\parbox{\columnwidth}{
    Network and Distributed System Security (NDSS) Symposium 2025\\
    24 - 28 February 2025, San Diego, CA, USA\\
    ISBN 979-8-9894372-8-3\\
    https://dx.doi.org/10.14722/ndss.2025.24568\\
    www.ndss-symposium.org
}
\hspace{\columnsep}\makebox[\columnwidth]{}}

\maketitle

\begin{abstract}
Binary code search plays a crucial role in applications like software reuse detection, and vulnerability identification.
Currently, existing models are typically based on either internal code semantics or a combination of function call graphs (CG) and internal code semantics. 
However, these models have limitations. 
Internal code semantic models only consider the semantics within the function, ignoring the inter-function semantics, making it difficult to handle situations such as function inlining.
The combination of CG and internal code semantics is insufficient for addressing complex real-world scenarios.
To address these limitations, we propose \binenhance{}, a novel framework designed to leverage the inter-function semantics to enhance the expression of internal code semantics for binary code search.
Specifically, \binenhance{} constructs an External Environment Semantic Graph (EESG), which establishes a stable and analogous external environment for homologous functions by using different inter-function semantic relations (\textit{e.g.}, \textit{call}, \textit{location}, \textit{data-co-use}).
After the construction of EESG, we utilize the embeddings generated by existing internal code semantic models to initialize EESG nodes. 
Finally, we design a Semantic Enhancement Model (SEM) that uses Relational Graph Convolutional Networks (RGCNs) and a residual block to learn valuable external semantics on the EESG for generating the enhanced semantics embedding. 
In addition, \binenhance{} utilizes data feature similarity to refine the cosine similarity of semantic embeddings.
We conduct experiments under six different tasks 
(\textit{e.g.}, under \textit{function inlining} scenario) 
and the results illustrate the performance and robustness of \binenhance{}. 
The application of \binenhance{} to HermesSim, Asm2vec, TREX, Gemini, and Asteria on two public datasets results in an improvement of Mean Average Precision (MAP) from 53.6\% to 69.7\%. Moreover, the efficiency increases fourfold.
\end{abstract}


\section{Introduction}\label{1}
Software development often reuses open-source code to reduce costs. 
However, this trend inadvertently propagates vulnerabilities of open-source code into billions of software systems~\cite{Cve-cve-2014-0106}~\cite{Cve-cve-2021-44228}, including industrial software and firmware.
The extensive workload of code auditing and the complexity of recursive code reuse results in substantial delays in vulnerability patching, with some systems experiencing an average delay of 352 days \cite{dong2023libvdiff}.
In 2023, a review~\cite{Synopsys} conducted by Synopsys on 1703 software projects exposed a notable problem: 96\% of these projects utilize open-source code, and 84\% of them have at least one known vulnerability.
In reaction to this widespread security risk, the emergence of binary code search has been proven to be a powerful method for automating the detection of insecure software components~\cite{luovulhawk}. 
This facilitates the prompt patching of identified vulnerabilities.

Binary code search entails the meticulous analysis of numerous binary codes to identify the most similar ones.
Hence, its applications span a wide range, including software reuse detection \cite{tang2022libdb,basit2005detecting,luo2014semantics}, vulnerability search \cite{luovulhawk,lin2021enbindiff,2018BINARM,2018VulSeeker}, firmware security analysis \cite{yang2021asteria,zhao2022large}, and patch presence testing \cite{xu2017spain,kargen2017towards,huang2017binsequence}.
Developing a general and effective binary code search solution is extremely challenging.
The syntactic structure of binary code can vary dramatically due to different compiler settings, such as optimization options.
Additionally, binary codes with similar syntactic structures may have different semantics~\cite{he2024code}. 
Therefore, a comprehensive understanding of the semantics of binary code is key to addressing this task.
With the rapid development of deep learning technology, current solutions for binary code search are to convert binary code into embeddings and obtain the similarity of binary code by calculating the similarity between embeddings. 
Based on the definition of Section~\ref{defintion}, we classify them into two categories: \textbf{internal code semantic models} and \textbf{external environment semantic models}.

Current solutions predominantly focus on learning the internal code features of functions. 
Gemini~\cite{xu2017neural} and Asm2vec~\cite{ding2019asm2vec} employ neural networks to perform semantic encoding within program structures, specifically control flow graphs. 
Similarly, TREX \cite{pei2020trex}, Palmtree \cite{li2021palmtree}, and others~\cite{wang2022jtrans}~\cite{yu2020order} utilize Transformer~\cite{vaswani2017attention} models for assembly code representation. 
Other studies, such as HermesSim~\cite{he2024code}, Asteria \cite{yang2021asteria}, VulHawk~\cite{luovulhawk} and XLIR \cite{gui2022cross}, 
utilize deep neural networks to convert intermediate representations, such as pseudocode and toy IR, into embeddings.
Furthermore, recent solutions integrate function call relationships with internal code features.  
$\alpha$diff \cite{liu2018alphadiff} and BMM \cite{guo2022exploring} utilize function call graph to aid in calculating similarity. 
Despite notable performance improvements in binary code search, these solutions have exhibited some limitations (example in Section~\ref{motivation_example}), particularly as the number of binary codes requiring comparison increases. 

\begin{figure*}
    \centering
    \includegraphics[width=\linewidth]{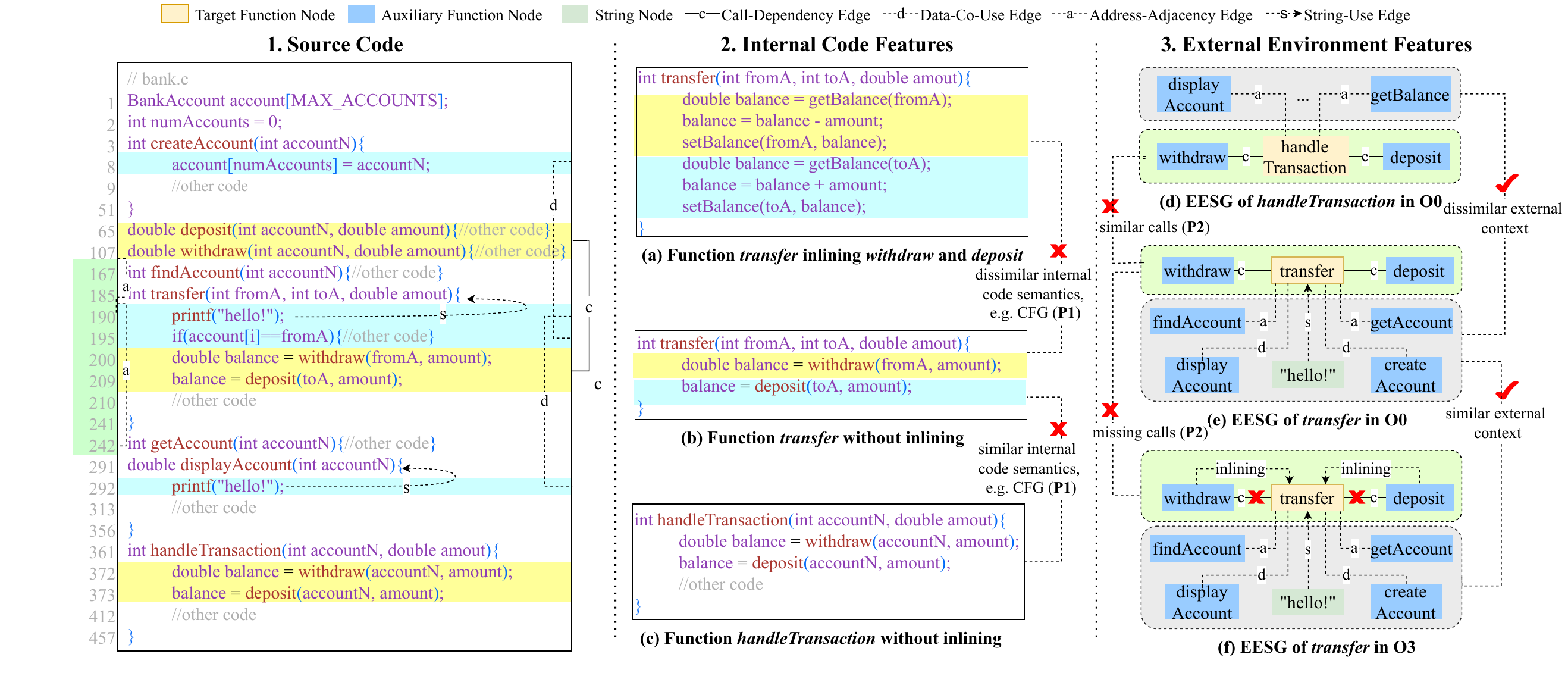}
    \caption{A Motivation Example of external environment semantics.}
    \label{fig:motivation}
\end{figure*}

Firstly, the internal code semantics of functions may exhibit substantial variations due to different compilation settings, encompassing factors like function inlining and splitting (identified as \textbf{P1}).
Function inlining, an optimization strategy during compilation, introduces the actual codes of other functions called by the function at the corresponding address (as shown in Figure~\ref{fig:motivation}.(a)). 
Function splitting, conversely, divides a single function into multiple sub-functions. Study~\cite{jia20221} demonstrates that compiler-caused function inlining can reach up to 70\%, leading to a significant decrease in the performance of binary code search. 
Moreover, the selective inlining method~\cite{chandramohan2016bingo} is lack of flexibility. 
This limitation underscores the inadequacy of relying solely on internal code semantics while overlooking the significance of external environment semantics.

Secondly, exclusive reliance on function call graphs (CG) for assistance is insufficient for addressing complex real-world scenarios (\textbf{P2}).
Relying solely on the external environment semantics of function calls introduces two primary issues that impact the effectiveness of binary code search: missing calls and similar calls. 
Missing calls occur when function call dependencies are absent, potentially eliminated during code reuse or optimization processes, such as strategies like function inlining.
Similar calls, meanwhile, involve similar function call relationships between non-homologous functions, leading to imprecise similarity assessments and false positives. 

Thirdly, current solutions exhibit limited scalability and struggle to cope with large-scale function search tasks (\textbf{P3}). 
The most recent Transformer-based methods, such as TREX~\cite{pei2020trex}, typically generate function embeddings with a dimension of 768. 
However, real-world scenarios often involve millions of functions, necessitating rapid comparison speeds. 
Our experiments (in Section~\ref{rq5}) revealed that in a pool of 10,000 functions, the time required to recall all the homologous functions of a binary file containing around 2500 functions is around 22 minutes using TREX's 768-dimension embeddings. 
The time increases exponentially with the growth of the function pool and the binary file size. In contrast, 128-dimension embeddings only require 5 minutes, emphasizing the efficiency challenges in large-scale function search. Moreover, simply reducing the dimensionality for retraining usually leads to substantial performance loss~\cite{li2021palmtree}.

Our central insight focuses on comprehensively utilizing external environment semantics (definition in Section~\ref{defintion}) to enhance the internal code semantic models.
To achieve it, several challenges must be addressed.  
First, it is crucial to carefully select environment features that can serve external environment semantics for a function within a complex environment. 
These selected features should positively contribute to the performance of binary code search tasks and exhibit robustness. 
Second, the integration of various external environment features, each representing distinct structures and semantic spaces, presents a challenge.
Determining an effective method for combining these diverse features to improve search tasks is crucial. 
Finally, an approach needs to be devised to appropriately apply the external environment semantics of the function, significantly enhancing its internal code semantics.

\begin{figure}
    \centering
    \includegraphics[width=\linewidth]{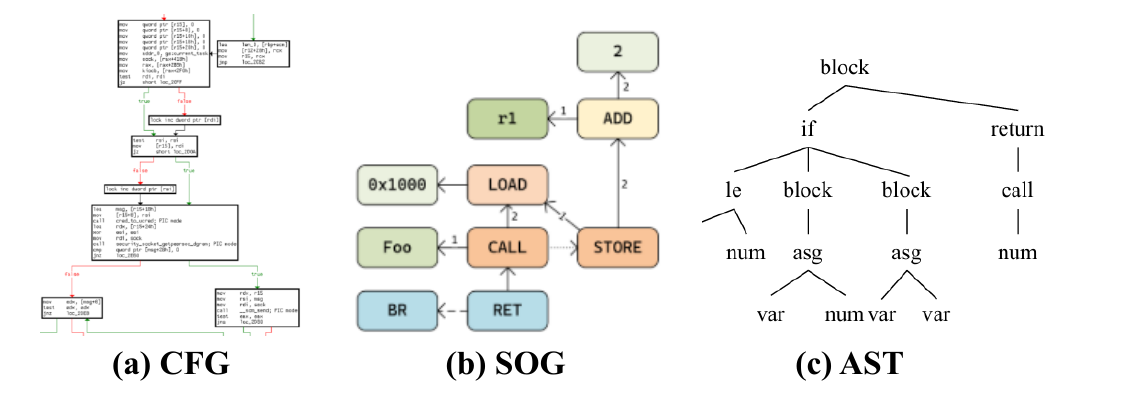}
    \caption{Examples of Internal Code Feature}
    \label{fig:internal}
\end{figure}

In this paper, we introduce \binenhance{}, a framework based on external environment semantics, specifically designed to tackle the aforementioned challenges. 
\binenhance{} differentiates itself from prior approaches by not solely relying on function call graph (\textbf{P2}). 
Instead, we have designed a novel External Environment Semantic Graph (EESG), which constructs four edges ($CD$, $DCU$, $AA$, and $SU$) among functions to model a function's external environment since these semantic edges are robust among different scenarios when they coexist. 
For \textbf{P3}, we employ whitening transformations~\cite{su2021whitening} to reduce the dimension of node initial embeddings generated by existing internal code semantics models~\cite{he2024code}~\cite{yang2021asteria}~\cite{xu2017neural}~\cite{pei2020trex}~\cite{ding2019asm2vec} and MPNET~\cite{song2020mpnet}, improving search task efficiency.
To address \textbf{P1}, after initializing the nodes of EESG, \binenhance{} proposes a Semantic Enhancement Model (SEM), which introduces external environment semantics into internal code semantic embedding. 
SEM utilizes Relational Graph Convolutional Networks (RGCNs)~\cite{schlichtkrull2018modeling} for updating external environment semantic embeddings in EESG and merges the dual embeddings of the function via a residual block to obtain an enhanced semantic embedding. 
Lastly, \binenhance{} includes a similarity combination module, leveraging data feature similarity to fine-tune semantic embedding similarity.
In summary, we have made the following contributions:
\begin{itemize}[leftmargin=10pt]
\item \textbf{Perspective.} 
This paper presents a novel perspective on binary code search by incorporating external environment semantics into internal embeddings. 
To the best of our knowledge, it is also the first work to demonstrate that this integration can significantly improve the identification of homologous functions. 
Our approach enriches the semantics of internal embeddings and addresses the problem of false filtering of those approaches based on function calls.
Additionally, it can be easily extended to other internal code semantic methods.
Moreover, this paper demonstrates the promising potential of designing new external features to enhance internal representations, rather than solely focusing on developing new features for internal code.
\item \textbf{Technique.} We implement \binenhance{}\footnote{https://github.com/wang-yongpan/BinEnhance}, a general framework that proposes four novel external semantic edges of our EESG to model stable external contexts in complex external environments and introduce an SEM to incorporate external semantics into internal embeddings to improve its representation ability.
\item \textbf{Study.} We conduct comprehensive experiments to evaluate the effectiveness of the \binenhance{}, including different function pool sizes, cross-architectures, and cross-optimization options, as well as scenarios such as function inlining and real-world vulnerability detection. Our results show that all existing binary code search methods overlook the importance of external environment semantics, leading to suboptimal performance.
\end{itemize}

\section{PRELIMINARIES}
\subsection{Definition}\label{defintion}
In this section, we introduce the terms applied in this paper. In the field of binary code search, \textbf{\textit{homologous functions}} refer to the functions compiled from the same source code with various compilation settings. The set of functions with homologous functions and non-homologous functions is termed a \textbf{\textit{function pool}}. Furthermore, this paper categorizes binary code search methods into two types: \textbf{Internal Code Semantic model} and \textbf{External Environment Semantic model}.

\begin{itemize}[leftmargin=10pt]
    \item \textbf{Internal Code Semantics.} It focuses on the target function itself and is derived from both the binary code embedded within the target function or from its derivatives, such as Control Flow Graphs (CFG) and Data Flow Graphs (DFG), which are constructed through code analysis of the binary function code. Figure~\ref{fig:internal} shows some examples of internal code features to represent internal code semantics.
    \item \textbf{External Environment Semantics.} It focuses on inter-function and is inferred from other supplementary functions (in the code segment) or any data (in the data segment) pertaining to the target function within the same binary, such as the function call relationships. In this paper, as shown in Figure~\ref{fig:motivation}, we present Call-Dependency, Data-Co-Use, Address-Adjacency, and String-Use as novel external environment semantics in our work (in Section~\ref{motivation_example} and~\ref{3.2.1}).
\end{itemize}

\subsection{Motivation Example}\label{motivation_example}
We elucidate our motivation with an example depicted in Figure~\ref{fig:motivation}.
This figure showcases how the external semantic exhibits superior performance than solely internal semantics in binary code similarity detection tasks.
It indicates the variance compiled from the same source code (\textbf{\textit{bank.c}}) under different compilation settings (\textbf{\textit{O0}} and \textbf{\textit{O3}}). 

\begin{itemize}[leftmargin=10pt]
\item \textbf{False Negative.} 
Figure~\ref{fig:motivation}.(a) and Figure~\ref{fig:motivation}.(b) illustrate the potential for false negative search results caused by function inlining during compilation.
In the figure, the \textbf{\textit{transfer}} function compiled under \textbf{\textit{O3}} incorporates the \textbf{\textit{withdraw}} and \textbf{\textit{deposit}} functions, through inlining, significantly altering its internal code semantics (\textbf{P1}). In contrast, Figure~\ref{fig:motivation}.(b) shows the result of the code compiled under \textbf{\textit{O0}} setting. There is no aforementioned function inlining, the dissimilar internal code semantics indicating that relying solely on it would fail to accurately match \textbf{\textit{transfer}} function compiled under \textbf{\textit{O3}} with its version compiled under \textbf{\textit{O0}} setting. Additionally, as depicted in Figure~\ref{fig:motivation}.(e) and Figure~\ref{fig:motivation}.(f), they highlight the problem (\textbf{missing calls in P2}), where \textbf{\textit{transfer}} loses call dependency due to function inlining, thus undermining the reliability of call-based approaches. 

\item \textbf{False Positive.}
Figures~\ref{fig:motivation}.(d) and \ref{fig:motivation}.(e) demonstrate the potential for false positive search results induced by similar function structures.
In these two figures, the \textbf{\textit{handleTransaction}} and \textbf{\textit{transfer}} functions, both highlighted in green, call \textbf{\textit{withdraw}} and \textbf{\textit{deposit}} functions, but they are not homologous functions. 
This scenario exemplifies a potential challenge (\textbf{similar calls in P2}), where solely using call dependencies might erroneously associate \textbf{\textit{transfer}} under \textbf{\textit{O0}} with \textbf{\textit{handleTransaction}} under \textbf{\textit{O0}} as homologous functions, impacting the accuracy of internal code semantics. Moreover, Figure~\ref{fig:motivation}.(b) and Figure~\ref{fig:motivation}.(c) indicates the similar internal code semantics of the \textbf{\textit{handleTransaction}} and \textbf{\textit{transfer}} functions, resulting in false positive.
\end{itemize}

To tackle these problems, we designed EESG to establish the external environment of the function. EESG is characterized by four external semantic edges and two types of nodes (in Section~\ref{3.2.1}). It ensures that even after compilation optimizations like function inlining, the EESG of \textbf{\textit{transfer}} remains largely unchanged (shown in grey in Figure~\ref{fig:motivation}.(e) and~\ref{fig:motivation}.(f)) and significantly different from the EESG of \textbf{\textit{handleTransaction}} (shown in grey in Figure~\ref{fig:motivation}.(d) and~\ref{fig:motivation}.(e)). This stability is crucial for effectively recalling homologous functions while minimizing interference from non-homologous ones.

\section{Design of \binenhance{}}

\subsection{Overview}

\begin{figure*}
    \centering
    \includegraphics[scale=.448]{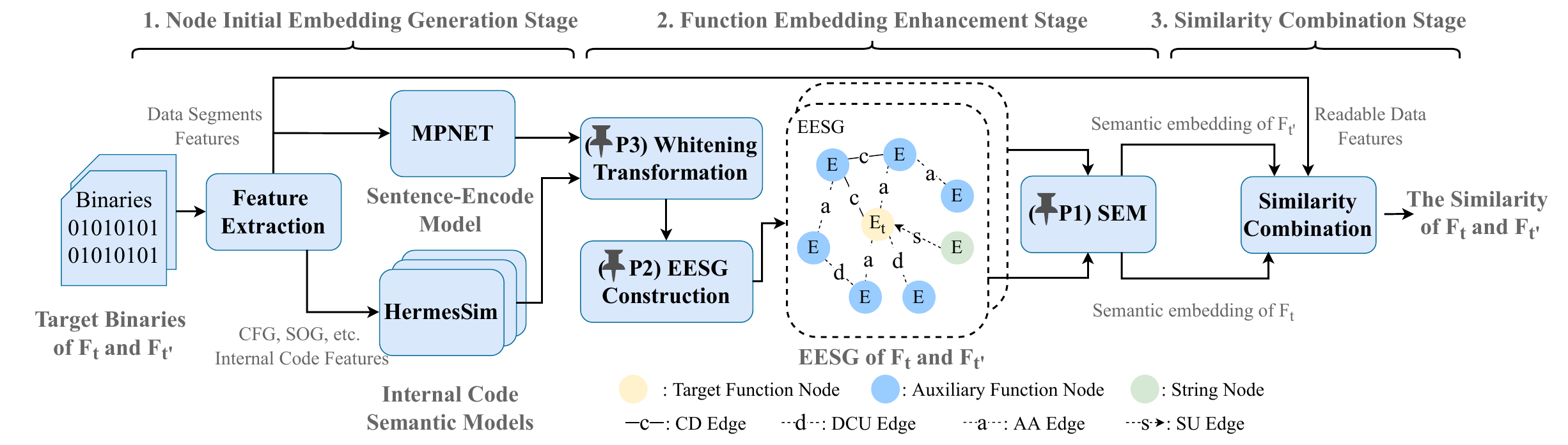}
    \caption{Workflow of \binenhance{}. \binenhance{} aims to calculate the similarity of functions $F_{t}$ and $F_{t'}$.}
    \label{fig:workflow}
\end{figure*}

\binenhance~seeks to utilize external function context and data features to assist in identifying homologous functions, 
especially when target functions lack clearly distinguishing internal code features.
To achieve this, we propose a robust external environment to aid in the identification of homologous functions. 
The enhancement provided by the designed external environment is pivotal for enhancing the recall performance of homologous functions.

Figure~\ref{fig:workflow} illustrates the workflow of \binenhance{}, comprising three stages: \textit{Node Initial Embedding Generation}, \textit{Function Embedding Enhancement}, and \textit{Similarity Combination}.
In the Node Initial Embedding Generation Stage, \binenhance{} employs the internal code semantic models (\textit{e.g.}, HermesSim~\cite{he2024code}) to generate the initial embeddings of functions, which are utilized for initializing nodes in the second stage. 
Specifically, we utilize official code to extract the internal code features and train the embedding model for generating all the function embeddings.
In addition, we extract the readable data features from the data segments of binaries to refine the similarity score post the second stage and utilize MPNET~\cite{song2020mpnet} to encode strings.
In the Function Embedding Enhancement stage, \binenhance{} constructs an External Environment Semantic Graph (EESG) acting as a supplementary source for the internal code semantics of functions. 
Subsequently, the embedding of string and function nodes are harmonized using Whitening Transformation~\cite{su2021whitening} to ensure the uniformity of embedding dimension in the encoded EESG. 
Finally, the encoded EESGs of the target functions are updated by SEM (Semantic Enhancement Model), which consists of multi-layer RGCN~\cite{schlichtkrull2018modeling} and a residual block, to enhance their nodes embedding.
In the Similarity Combination stage, \binenhance{} computes the Jaccard similarity of the data features and utilizes this score to adjust the cosine similarity output from stage two.
This computation is implemented through a feed-forward neural network. 
The resulting similarity score is then considered as the final measure of similarity between the two input functions.

\subsection{Node Initial Embedding Generation}

In this section, we introduce \binenhance{} how to utilize the function embeddings generated by HermesSim and other internal code semantic models, as well as how to initialize the nodes embedding of EESG.

Section~\ref{3.2.1} introduces the two types of nodes in the EESG: function nodes and string nodes, necessitating distinct methods for embedding generation.
We observe that existing internal code semantic models, such as HermesSim~\cite{he2024code}, are well-suited for function node embeddings. 
Additionally, HermesSim can be seamlessly replaced with other more advanced models as needed.
In this paper, we select five existing internal code semantic models to generate function embeddings and respectively enhance their performance using \binenhance{} (in Section~\ref{baselines}).
Specifically, we utilize the official code provided by each model to extract the required internal features, such as the control flow graph (CFG) by Gemini~\cite{xu2017neural}, the abstract syntax tree (AST) used by Asteria~\cite{yang2021asteria}, and the semantic-oriented graph (SOG) used by HermesSim~\cite{he2024code}, and so on.
Following the extraction process, we retrain the models using the official training codes and hyperparameters to prevent performance discrepancies arising from varying dataset distributions.
Subsequently, we employ the retrained internal embedding models to generate function embeddings within the target binary. 

For string nodes, we utilize the widely recognized MPNET~\cite{song2020mpnet}, a sentence-transformer and pre-trained English language model, to transform English strings into embeddings, ensuring rich linguistic representation.
In addition, based on the existing research~\cite{yuan2019b2sfinder}, there may be some consistent data features within the $.rdata, .data, .bss, .idata$, and other data segments of binary files.
In binary files readable strings predominantly reside in these segments or symbol tables,
and those residing in the symbol table are often stripped, thus prioritizing data segment strings for stability.
Global data, such as global constant arrays and integer constants, are shared across multiple functions.
Like OSLDetector~\cite{zhang2020osldetector}, we discover that most strings with less than 5 (e.g., $time$) are more likely to be reused, with lower specificity, and thus we filter them.
B2Sfinder~\cite{yuan2019b2sfinder} demonstrates that these features are helpful and typically not stripped off. We extract these data features for the third stage. While extracting each data feature, we concurrently save a list of addresses for all functions that use this data within the code segment.
This facilitates the subsequent construction of the EESG (in Section~\ref{3.2.1}).

\subsection{Function Embedding Enhancement}

In this section, we introduce the details construction process of EESG and how to use SEM to learn the external semantics of EESG.

\binenhance{} employs EESG to model the semantic information of the functions' external contexts.
This is in contrast to previous approaches that relies solely on function call relationships within the function's call graph (CG). 
After generating the initial embeddings of the two types of nodes on EESG, two problems arise from different initialization methods.
Firstly, there's a discrepancy in the embedding dimensions between string nodes and function nodes (the former is fixed at $768$ dimensions by MPNET, while the latter varies depending on the internal code semantic models applied).
Secondly, high-dimensional embeddings, such as $768$ dimensions, impose significant computational and memory burdens, especially in extensive binary function search tasks.

\subsubsection{Whitening Transformation}\label{wt}

\begin{algorithm}
    \setstretch{1.0}
        \SetAlgoLined 
	\caption{Whitening Transformation Method}
	\KwIn{Node Embeddings $X$, Target Dimension $d_{t}$}
	\KwOut{Whitening Embeddings $E$}
    \label{algo:a2}
	\BlankLine
        \SetKwProg{Def}{Function}{:}{end}
        \SetKwFunction{}{}
        \Def{Whitening\_Transformation($X$, $d_{t}$)}{
        \SetAlgoVlined
            $\mu = \displaystyle \frac{1}{n}  {\textstyle \sum_{i=1}^{n}}X_{i} $\;
            $Cov= \displaystyle \frac{1}{n-1}  {\textstyle \sum_{i=1}^{n}}\left ( X_{i} -\mu  \right ) \left ( X_{i}-\mu   \right ) ^{T}$\;  
            $\displaystyle V, D, V^{T} = SVD\left ( Cov \right ) $\;
            $\displaystyle W=VD^{-\frac{1}{2} } V^{T} \left [ :, \space\space\space\space:\space\space\space\space d_{t} \right ] $\;
            \For{$i=1$ to $n$}
            {
                $E_{i} =\displaystyle \left ( X_{i} - \mu   \right ) W$\;
                $E_{i} =\displaystyle \frac{E_{i} }{\sqrt{ {\textstyle \sum_{j=1}^{d_{t}}} E_{i,j}^{2}  } } $\;
            }
            return $E$\;
            }
\end{algorithm}

To address these problems, we apply the whitening transformation~\cite{su2021whitening} for dimensionality reduction, enabling two nodes to encode into a unified vector space. 
In Algorithm \ref{algo:a2}, 
we elaborate on the details of applying the whitening transformation method for linear dimensionality reduction.
Specifically, the input $X \in \mathbb{R}^{n \times d}$ represents the node embedding matrix of the EESG, where $n$ is the total number of nodes and $d$ is the initial dimension of embeddings.
The algorithm's output, $E\in \mathbb{R}^{n \times d_{t}}$, is the transformed node embedding after whitening, where $d_{t}$ signifies the reduced dimension of nodes. 
The algorithm initiates by calculating the mean and covariance matrix of $X$ (lines 2 and 3). Subsequently, an eigenvalue decomposition of the covariance matrix is executed using Singular Value Decomposition ($SVD$) (line 4). Here, $D$ is defined as a diagonal matrix containing the eigenvalues, and $V$ is an orthogonal matrix with its columns representing the corresponding eigenvectors. The whitening transformation is then facilitated through the construction of a whitening matrix $W$, derived from $V$ and $D$ (line 5). This matrix is applied to the initial embeddings to yield the whitening embeddings (line 7). 
In the final step, normalization of the whitening embeddings is achieved by first calculating the $L2$ norm for each embedding $E_{i}$, followed by division of each embedding by its respective $L2$ norm to complete normalization. 
This results in a set of standardized, dimensionally-reduced node embeddings suitable for further semantic learning. 

\subsubsection{Details of EESG and its Construction}\label{3.2.1}
EESG construction focuses on constructing knowledge from two distinct perspectives: nodes and edges. Specifically, EESG is represented as a directed graph, denoted as $G=(N, E, N_{e}, E_{t})$, where $N$ represents the set of different types of nodes, $E$ $\subseteq$ $N \times N$ is the set of edges, $N_{e}$ is the set of node embeddings, and $E_{t}$ is the set of edge types.

\noindent\textbf{Nodes.} In the EESG, each node ($N_{i}$) represents one of two distinct entities: \ding{182} \textbf{Functions within code segments} and \ding{183} \textbf{Readable strings within data segments}. 
Function nodes are further categorized into two types: the target function ($F_{t}$) and the auxiliary function ($F_{a}$).
The auxiliary function $F_{a}$ refers to all functions that have a direct or indirect semantic relationship with the target function.
The purpose is to construct complete semantic information about the external environment of the target function ($F_{t}$).
Readable string nodes, extracted from the data segment, represent strings used in function nodes. 
After the whitening transformation, the embedding dimensions of the two types of nodes become consistent, preparing them for subsequent SEM to update the embedding of nodes.

\noindent\textbf{Edges.} 
Edges express not only the semantic information of target and auxiliary nodes but also the function nodes and readable string nodes.
Building upon existing research \cite{karande2018bcd}, we categorize edges within the EESG into the following four distinct types.
\noindent \ding{182} \textbf{Call-Dependency ($CD$) edge}: Within the x86 instruction set, a function $F_{i}$ calling another function $F_{j}$ via the $CALL$ instruction denotes a $CD$ edge.
For instance, considering a compiler adopting function inlining strategy, a function $F_{j}$ is inlined into another function $F_{i}$, resulting in a modified function $F_{i}'$ of $F_i$.
From the view of static analysis, $F_{i}'$ exhibits substantial deviations in semantic content compared to the original $F_{i}$. To mitigate this, establishing an edge between $F_{i}$ and $F_{j}$ is proposed. 
This connection allows $F_{i}$ to supplement and access $F_{j}$'s semantic content, thereby aligning $F_{i}$ more closely with the modified semantic context of $F_{i}'$. 
Nevertheless, relying exclusively on $CD$ edges presents certain constraints (in Section~\ref{1}).
To address these, three additional types of edges are introduced below, designed to enrich the semantic representation and alleviate the impact introduced by optimization strategies, including inlining.
\noindent \ding{183} \textbf{Data-Co-Use ($DCU$) edge}: Binary files typically contain data in segments like $.data$, $.bss$, $.rodata$, and $.idata$.
We focus on constant readable strings and global data within these segments. 
We create bidirectional $DCU$ edges of string or global data by linking two functions using the same constant readable string or global data.
These edges introduce data semantics into function semantics and counteract false positives due to structural similarities introduced by $CD$ edges in the same call graph.
Notably, even when the symbol names of global data are stripped, we can establish these edges based on data addresses. 
\noindent \ding{184} \textbf{Address-Adjacency ($AA$) edge}: Inspired by BinaryAI~\cite{jiang2024binaryai} and LibDX~\cite{tang2020libdx}, we observed that after the compiler compiles all source functions within the same source file, the order of function addresses in the binary file may differ from that in the source code due to different compilation settings. However, the relative locality of functions in the code segment remains unchanged. Therefore,
we establish a directional $AA$ edge based on the starting address and size of each function. Specifically, when the address of function $F_{i}$ exceeds that of $F_{j}$, an $AA$ edge is created from $F_{i}$ to $F_{j}$. 
This edge signifies that $F_{i}$ precedes $F_{j}$ in the address sequence. 
Conversely, an edge from $F_{j}$ to $F_{i}$ indicates that $F_{i}$ follows $F_{j}$. 
This methodology is particularly effective in capturing the address semantics of functions, a critical aspect of open-source code reuse.
Open-source functions are predominantly reused as modular units and are stored adjacently~\cite{yang2022modx, tang2020libdx}. 
This adjacency is preserved during compilation, allowing our $AA$ edges to represent the address semantics between functions.
Additionally, this approach addresses the challenge of enhancing semantic representation in scenarios where the above $CD$ and $DCU$ edges are insufficient or absent.
\noindent \ding{185} \textbf{String-Use ($SU$) edge}: 
Prior research~\cite{xu2017neural} indicates the stability of strings used in functions across different compilation environments.
Therefore, we establish edges from readable string nodes to function nodes, aiming to augment the ability of a function node to identify homologous nodes based on the semantic information in readable strings.

\begin{algorithm}
    \setstretch{0.8}
        \SetAlgoLined 
	\caption{EESG Construction Algorithm}
	\KwIn{Feature Sets $s_{e}$, Function Sets $s_{u}$, Target Function $f_{t}$ and Max Depth $md$}
	\KwOut{the EESG of $f_{t}$}
    \label{algo:a1}
	\BlankLine
        $eesg, ns, es \leftarrow NULL$\;
        \SetKwProg{Def}{Function}{:}{end}
        \SetKwFunction{}{}
        \Def{EESG\_Construct($s_{e}$, $s_{u}$, $f_{t}$, $md$)}{
        \SetAlgoVlined
            $nss \leftarrow f_{t}$\;
            $eesg.addNode(f_{t}, type=function)$\;
            $ns.add(f_{t})$\;
            \For{$d=0$ to $md$}
            {
                $nes \leftarrow NULL$\;
                \For{$n \in nss$}
                {
                    $cd, dcu, aa, su \leftarrow $ GetExternal($s_{u}$, $s_{e}$, $n$)\;
                    $nes.add(BuildEdges(n, cd, 0, 1, function))$\;
                    $nes.add(BuildEdges(n, aa, 2, 3, function))$\;
                    $nes.add(BuildEdges(n, dcu, 4, 4, function))$\;
                    $BuildBiEdges(n, su, 5, 5, string)$\;
                }
                $nss \leftarrow nes$
            }
            return $eesg$\;
            }
        \BlankLine
        \SetKwProg{Functionl}{Function}{:}{end}
        \Functionl{BuildEdges($src$, $des_{f}$, $r_{1}$, $r_{2}$, $node\_type$)}{
        \SetAlgoVlined
            $n \leftarrow NULL$\;
            \For{$des \in des_{f}$}
                {
                    $e \leftarrow (src, des, type=r_{1})$\;
                    \If{$des \not \in ns$}
                    {
                        $essg.addNode(des, type=node\_type)$\;
                        $n.add(des)$\;
                        $ns.add(des)$\;
                    }
                    \If{$e \not \in es$}{
                        \If{$node\_type == function$}{
                            $eesg.addEdge(src, des, type=r_{1})$\;
                            $es.add(src, des, type=r_{1})$\;
                        }
                        $eesg.addEdge(des, src, type=r_{2})$\;
                        $es.add(des, src, type=r_{2})$\;
                    }
                }
        return $n$\;
        }
\end{algorithm}

\noindent\textbf{EESG Construction Algorithm.}
EESG construction initiates from the function node $F_{t}$ and is gradually completed recursively by obtaining the auxiliary function nodes $F_{a}$ based on four types of edges. 
As detailed in Algorithm~\ref{algo:a1}, the process begins with the initialization of the EESG, denoted as $eesg$. 
Because of the existing multiple identical edges between two nodes (\textit{e.g.}, multiple calls), this step is followed by creating two lists: $ns$ for minimizing redundant nodes, and $es$ for eliminating duplicate edges.
Considering that the auxiliary functions staying far away from the function node $F_{t}$ may have a weaker effect on its semantic enhancement, we set a max depth for recursive execution and to limit the size of $eesg$, marked as $md$. 
Empirical evidence from our experiments suggests that a maximum depth of $4$ strikes an optimal balance.
Algorithm~\ref{algo:a1} systematically constructs the EESG, layer by layer, commencing with the function node $F_{t}$. 
It identifies non-repetitive function nodes in successive layers, represented as $nes$ for the next layer and $nss$ for the current one.
By traversing each node in $nss$, we ascertain the subsequent layer's function nodes, $nes$, through three edge types: call-dependency ($cd$), data-co-use ($dcu$), and address-adjacency ($aa$).
The algorithm establishes bi-directional edges between function nodes, with $0$/$1$ indicating call/be-called dependency edges, $2$/$3$ for pre/post address adjacency edges, and $4$ for data-co-use edges. 
Furthermore, the algorithm constructs string nodes and the edges linking these nodes to their corresponding function nodes. 
Based on the string content $su$ utilized by the function node, the edges are categorized as type $5$.
The process culminates with the algorithm returning the $eesg$ of the $F_{t}$.
The function $GetExternal$ at line 9 retrieves all function nodes and string nodes connected with the function node $n$ through the four types of edges.

\subsubsection{Semantic Enhancement Model}

After the construction of the EESG for the target function $F_{t}$, we observe that the inherent ability of graph neural networks (GNNs)~\cite{schlichtkrull2018modeling} to aggregate neighboring node information for updating target node embedding aligns well with the requirement to learn the semantic knowledge of the external environment of $F_{t}$ node.
Therefore, we designed the SEM model based on GNN to enhance the representation of function nodes, capable of learning the structure of the EESG and incorporating both external and internal semantics.

\begin{figure}
    \centering
    \includegraphics[scale=.37]{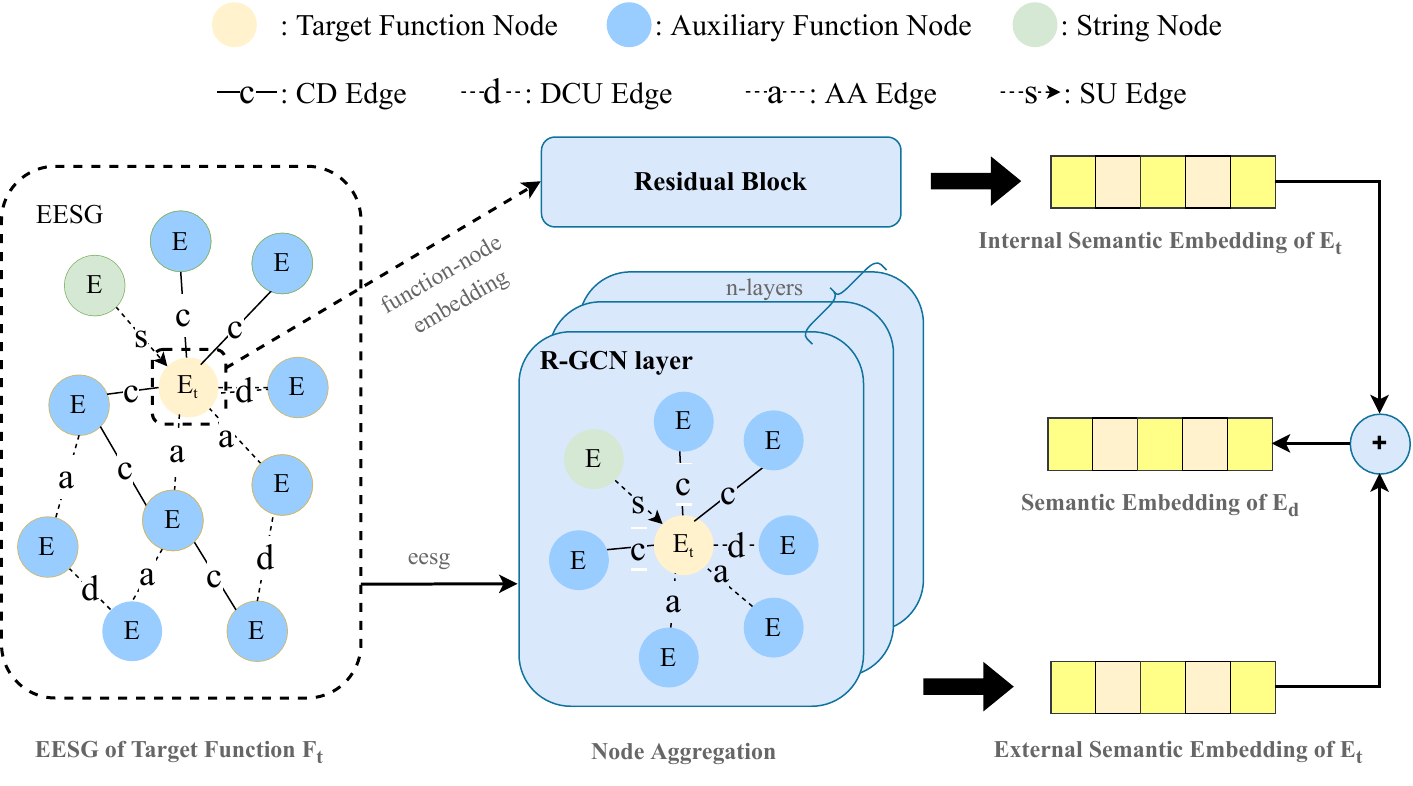}
    \vspace{-1.5em}
    \caption{Structure of SEM}
    \label{fig:eeslm}
    \vspace{-2.5em}
\end{figure}

Figure~\ref{fig:eeslm} depicts the structure of SEM. 
After encoding the EESG, we select a GNN to integrate the embedding of neighboring nodes relevant to the $F_t$.
In particular, we select Relational Graph Convolutional Neural Network (RGCN)~\cite{schlichtkrull2018modeling}. 
Notably, while utilizing a more advanced GNN could enhence SEM's performance, it lies outside of this study's primary focus. 
Compared with the GCN~\cite{kipf2016semi}, the RGCN contemplates node interrelations and adeptly addresses different edge relationship impacts within the graph structure, aligning seamlessly with EESG's graph structure.
\begin{equation}
    E_{i}^{(l+1)}=  LeakyRelu(\sum_{r\in R}^{} \sum_{j\in N_{i}^{r} }^{} \frac{1}{c_{i,r}} W_{r}^{(l)}h_{j}^{(l)})
    \label{exp:exp1}
\end{equation}
\begin{equation}
    E_{t} = RGCN(F_{t} ) +  Resisual\_block(HermesSim(F_{t} ))
    \label{exp:exp2}
\end{equation}

For the edges of EESG, we set different trainable parameters for different edge types by the Pytorch~\cite{pytorch} framework based on the RGCN model.
It guides function nodes to learn unique semantic information based on edge types. Specifically, for each function node $F_{i}$, we update its function embedding following Equation~\ref{exp:exp1} (from the RGCN model), where $N_{i}^{r}$ denotes the set of neighbor nodes related to $F_{i}$ through relation $r$, while $C_{i,r}$ represents a normalization factor set to the cardinality of $N_{i}^{r}$. $W_{r}^{\left ( l \right ) }$ denotes the linear transformation function, applied uniformly across neighbor nodes connected by identical edge types, with the count of $W_{r}^{\left ( l \right ) }$ corresponding to the number of EESG's edge types. The layer count, $l$, indicates the layer number of the RGCN network.
SEM's structure facilitates the embedding amalgamation of all neighboring nodes for a given function node at each layer through adjustable parameters, continuously refining its embeddings.
As the number of neural network layers increases, each target function node incrementally learns the semantic information from increasingly distant auxiliary function nodes.
Since we adopt four levels of EESG in depth (in Section \ref{3.2.1}), consistently, a four-layer neural network structure for SEM is applied to learn the semantics of all the auxiliary function nodes of the target node on EESG.
After these neural network layers are updated, each function node's embedding is encoded with the semantics of its comprehensive external environment. Thus, $F_{t}$ learns its external semantic embedding.
To better integrate the internal semantics of the function with its external semantics and prevent it from forgetting its internal semantics, SEM introduces a residual block with trainable parameters. 
This residual block transforms the initial embedding of $F_{t}$ into an internal semantic embedding with a dimension consistent with the dimension of the external semantic embedding by the linear layer of pytorch.
Finally, the SEM integrates two embeddings, internal semantic embedding, and external semantic embedding, to obtain the final semantic embedding representation of the target function $F_{t}$, as shown in Equation~\ref{exp:exp2}.

\subsection{Similarity Combination}
In this section, we describe how to combine data feature similarity with the cosine similarity of embeddings and outline the training and inference process for the entire framework.

\subsubsection{Jaccard similarity}
We combine the similarities of data features between homologous functions with the cosine similarity of semantic embeddings by a feed-forward neural network (FFN) to obtain the final similarity.
While prior studies \cite{yuan2019b2sfinder} have utilized these data features to identify binary functions, their efficacy in recognizing homologous functions is limited.
Firstly, not all functions contain such data features, as evidenced by the study~\cite{dong2023libvdiff}. 
Secondly, as discussed in Section \ref{1}, due to the existence of optimization strategies such as function inlining, which cannot be resolved by data features. Therefore, These features serve as an auxiliary tool to fine-tune the calibration of semantic embedding similarity, rather than being the sole basis of function recall. 

We employ the Jaccard similarity to calculate the similarity of data between two functions.
In particular, this similarity is computed as $Sim = \displaystyle \frac{\left | F_{m} \cap F_{n}\right | }{\left |F_{m}\cup F_{n}  \right | }$, where $\cap / \cup$ represents the intersection/union of the features of the functions $F_{m}$ and $F_{n}$.
Next, we compute the cosine similarity ($Sim_{cos}$) between the semantic embedding representations ($E_{d}$ and $E_{d'}$) of the two functions, as derived from the SEM.
As shown in Equation~\ref{exp:exp3}, this cosine similarity is then concatenated with the data similarities to construct a similarity vector.
Finally, an FFN is utilized to predict the ultimate similarity between the two functions by the similarity vector.
It's noteworthy that even in the absence of data features for similarity combination, our SEM (without data features) demonstrates commendable performance in semantic embedding, as evidenced by the results in the ablation study (in Section~\ref{ablation}).
\begin{equation}
\vspace{-0.4em}
    Sim=  Tanh(FFN(Concat(Sim_{cos}, Sim_{data})) 
    \label{exp:exp3}
\end{equation}
\subsubsection{Framework Training and Inference}

As shown in Equation \ref{exp:exp3}, the final similarity score generated by the framework is derived from the $Tanh$ activation function, yielding values within the range of $[-1, 1]$.
Therefore, we employ the Mean Squared Error (MSE) as our loss function, as delineated in Equation \ref{exp:exp4}, where $m$ denotes the batch size during the training process, while $Sim_{p}$ and $Sim_{n}$ represent the similarity scores of homologous and non-homologous functions to the target function, respectively.
The target similarity for homologous functions is set to $1$, and conversely, $-1$ for non-homologous functions.
Furthermore, we utilize the Adam optimizer to train the parameters, aiming to minimize the loss.
Meanwhile, in the \binenhance{} framework, only the parameters of the SEM and FFN in the Similarity Combination participate in the end-to-end training of the framework.
Other modules, such as HermesSim and MPNET, are only utilized for node initial embedding generation.
\begin{equation}
    Loss= \frac{1}{m} \sum_{m}^{}\frac{1}{2} (1-Sim_{p}^{}  )^{2} +  \frac{1}{2} (1+  Sim_{n})^{2} 
    \label{exp:exp4}
    \vspace{-0.4em}
\end{equation}

When calculating the similarity between functions, we follow the common practice in most binary code search methods by first extracting features and constructing the function's EESG. 
We then execute existing internal code semantics models and MPNET, followed by a whitening transformation to complete the node embedding assignment for the EESG. After updating the nodes with SEM, we combine the data feature similarities with embedding cosine similarity to obtain the final function similarity.

\section{EVALUATION}
In this section, we apply \binenhance{} to the existing state-of-the-art binary code search methods for the performance evaluation of multiple related tasks.
In the experiments, we aim to answer the following research questions: 
\textbf{RQ1}: How much could be improved when \binenhance{} is applied to baselines (including HermesSim~\cite{he2024code}, Asteria~\cite{yang2021asteria}, Asm2vec~\cite{ding2019asm2vec}, TREX~\cite{pei2020trex}, and Gemini~\cite{xu2017neural})?
\textbf{RQ2}: Is \binenhance{} robust against different compiler optimization options and architectures?
\textbf{RQ3}: Does \binenhance{} effectively alleviate the impact of function inlining in binary code search?
\textbf{RQ4}: What is the contribution of each part in \binenhance{}?
\textbf{RQ5}: Does \binenhance{} improve the efficiency of baselines?
\textbf{RQ6}: What is the performance of \binenhance{} in detecting 1-day vulnerabilities in real-world firmware?
\subsection{Implementation and Experiment Setup}
We use IDA Pro 7.5~\cite{IDApro} for disassembling binary files and Python 3.8 for programming \binenhance{} with libraries, such as DGL~\cite{dgl}, and Pytorch~\cite{pytorch}. 
Our evaluation is conducted on an Ubuntu 22.04 with an Intel Xeon 128 core 3.0GHz CPU, 1TB memory, and 2 Nvidia V100 32GB GPUs.
\begin{table*}
  \caption{The details of three public datasets and one firmware dataset}
  \vspace{-0.5em}
  \label{tab:dataset}
  \begin{tabular}{ccccccc}
    \toprule
    Dataset & architectures & options &  Functions &  Applications & Projects & Source \\
    \midrule
    D1 & ARM, X86, X64 &  O0-O3 &  2,751,667 & RQ1, RQ2 & 260 & Asteria\\
    D2\_norm & ARM, MIPS, X86 &  O0-O3 &  1,654,805 & RQ1, RQ2, RQ3, RQ4, RQ5 & 51 & BinKit\\
    D2\_noinline & ARM, MIPS, X86 &  O0-O3 &  1,991,864 & RQ3 & 51 & BinKit\\
    D3\_firmware & ARM, MIPS, X86 &  UNKNOWN &  6,817(binaries) & RQ6 & 37 & Real-world\\
  \bottomrule
\end{tabular}
\end{table*}
\begin{table}
  \centering
  \caption{Function Embedding dimensions}
  \vspace{-0.5em}
  \label{tab:dim}
  \begin{tabular}{l|cc}
    \toprule[2pt]
    Model & Original model & +\binenhance{} \\
    \cline{1-3}
    Gemini & 64 &  64\\
    \cline{1-3}
    Asm2vec & 200 &  64\\
    \cline{1-3}
    Asteria & 150 &  128\\
    \cline{1-3}
    TREX & 768 &  128\\
    \cline{1-3}
    HermesSim & 384 &  128\\
  \bottomrule[2pt]
\end{tabular}
\end{table}
\vspace{-1.5em}
\subsubsection{Datasets}
In the experiments, as shown in Table~\ref{tab:dataset}, we select three public datasets \cite{yang2021asteria, kim2022revisiting} and collect a firmware dataset. 
We divide the public datasets into a training set, validation set, and testing set in an 8:1:1 ratio based on the granularity of the project. 
The test data differs from the training and validation data to make the evaluation result reliable.
\vspace{-1.0em}
\begin{itemize}[leftmargin=10pt]
\item \textbf{Dataset $D1$}. The dataset is from $Asteria$~\cite{yang2021asteria}, including three architectures: $ARM$, $X86$, and $X64$ with four levels of optimizations: $O0$, $O1$, $O2$, and $O3$. In total, this dataset contains $2,751,667$ functions.
It includes widely utilized tool libraries like $binutils$ and spans various categories, such as communication (\textit{e.g.}, $openssl$, $acl$), compression (\textit{e.g.}, $p7zip$, $szip$), file operation (\textit{e.g.}, $file$, $json$) and others, covering $260$ distinct projects. 
\item \textbf{Dataset $D2\_norm$}. The dataset is from $BinKit$ \cite{kim2022revisiting}, $D2\_norm$ spans three architectures ($ARM$, $MIPS$, $X86$) and four optimization levels ($O0$, $O1$, $O2$, $O3$), featuring a total of $1,654,805$ functions from $51$ projects like $coreutils$ and $gnu$. 
$D2\_norm$ uses standard compilation.
\item \textbf{Dataset $D2\_noinline$}. $D2\_noinline$ and $D2\_norm$ contain identical software projects. The key distinction lies in their compilation process: $D2\_noinline$ prohibited function inlining during compilation. 
Therefore, $D2\_noinline$ contains $1,991,864$ functions, exceeding the size of $D2\_norm$.
\item \textbf{Dataset $D3\_firmware$}. It consists of $37$ firmware images from $8$ vendors (\textit{e.g.}, $DLINK$ and $ASUS$) that span a variety of device categories such as cameras and routers. 
We use $Binwalk$ \cite{binwalk.} to parse the images and extract the $6,817$ binaries from them. 
We collect them to assess the ability of \binenhance{} to detect 1-day real-world vulnerabilities.
\end{itemize}
\vspace{-1.0em}
\subsubsection{Evaluation Metrics}
In practice, the binary code search tasks are often accompanied by a big function pool where most are non-homologous functions. 
Therefore, we select Mean Average Precision ($MAP$) as the evaluation metric, as shown in Equation~\ref{exp:exp5}, where $AP$ denotes the average precision of one prediction result, and $k$ is the ranking position of the prediction functions. 
$P(k)$ is the proportion of homologous functions among the top-$k$ prediction functions.
$sim(k)$ indicates whether the $k_{th}$ prediction is a homologous function or not, specifically $1$ for yes, and $0$ for no. 
$N$ indicates the total number of homologous functions in this test data. 
$MAP$ represents the mean value of the average precision of the multiple prediction results.  
$q$ is the serial number of the prediction and $Q$ is the total number of predictions. $AP(q)$ is the average precision of the $q_th$ predicted result.
\begin{equation}
    AP=\frac{\sum_{k=1}^{N}(P(k)\times sim(k)) }{N },\thickspace
    MAP=\frac{\sum_{q=1}^{Q}AP(q) }{Q} 
    \label{exp:exp5}
\end{equation}

\subsubsection{Baselines}\label{baselines}
To demonstrate the effectiveness of \binenhance{} to existing methods, we select the following state-of-the-art (SOTA) methods of binary search from various perspectives as baselines based on their open-source availability and the requirement for generating function embeddings. 
For all the baselines, we utilize the official code and default parameters. 
Moreover, we retrain all the baselines for each dataset (except $D3\_firmware$) using the official training code to prevent performance discrepancies from varying dataset distributions.
\begin{itemize}[leftmargin=10pt]
\item \textbf{Gemini~\cite{xu2017neural}.} This baseline utilizes Structure2vec \cite{dai2016discriminative} to learn ACFG and encode function embeddings.
\item \textbf{Asm2vec~\cite{ding2019asm2vec}.} This baseline uses the PV-DM model to learn instruction sequence embedding representations.
\item \textbf{Asteria~\cite{yang2021asteria}.} This baseline utilizes the Tree-LSTM to learn ASTs and obtain embedding representations.
\item \textbf{TREX~\cite{pei2020trex}.} This baseline inputs the sequence generated by the micro-trace into the transformer model for training.
\item \textbf{HermesSim~\cite{he2024code}.} This baseline utilizes Ghidra~\cite{Ghidra} to lift binary code to the toy IR and proposes a Semantic-Oriented Graph (SOG) to generate function embeddings.
\end{itemize}

For external semantic learning methods, we also evaluate BMM~\cite{guo2022exploring} that use function call relationships (in Section~\ref{ablation}). 
We additionally evaluate the influence of varying dimensions on function embedding, aiming to identify the suitable dimension.
Ultimately, considering both the MAP score and computational time, 
we employ the whitening transformation module (in Section~\ref{wt}) to reduce the dimensions of the embedding derived from the baseline methods. 
Details of the dimension setting are shown in Table \ref{tab:dim}.

\subsection{RQ1: MAP Improvement Evaluation}

\begin{table*}[t]
\centering
\vspace{-0.5em}
\renewcommand{\arraystretch}{1.1}
\setlength {\tabcolsep} {2.5pt}
\caption{MAP scores of different methods of binary code search tasks in different function pool sizes.}
\vspace{-0.7em}
\begin{threeparttable}
\begin{tabular}{l|cccccc|cccccc}
\toprule[2pt]
\multicolumn{1}{c|}{\multirow{2}{*}{\textbf{Models}}} & 
\multicolumn{6}{c|}{\textbf{$D1$}} & 
\multicolumn{6}{c}{\textbf{$D2\_norm$}} \\

\cline{2-13}

\multicolumn{1}{c|}{} & \textbf{P=2} & \textbf{P=128} & \textbf{P=1024}&\textbf{P=4096} & \textbf{P=10000} & \textbf{Avg.} & \textbf{P=2} & \textbf{P=128} & \textbf{P=1024}&\textbf{P=4096} & \textbf{P=10000} & \textbf{Avg.}\\
\hline


\multirow{1}{*}{Gemini} &93.5&	60.0&	40.2&	28.1&	21.7&	48.7&	95.9&	64.3&	37.7&	24.0&	17.3&	47.8\\
\multirow{1}{*}{Gemini+\binenhance{}} &\textbf{95.4}&	\textbf{74.1}&	\textbf{59.4}&	\textbf{48.7}&	\textbf{42.1}&	\textbf{63.9}&	\textbf{96.1}&	\textbf{77.5}&	\textbf{63.6}&	\textbf{52.8}&	\textbf{45.1}&	\textbf{67.0}\\
\cline{1-13}

\multirow{1}{*}{TREX} &94.9&	65.4&	42.8&	28.8&	21.4&	50.7&	74.8&	20.8&	14.0&	11.2&	9.3&	26.0\\
\multirow{1}{*}{TREX+\binenhance{}} &\textbf{96.0}&	\textbf{75.2}&	\textbf{58.2}&	\textbf{45.6}&	\textbf{38.6}&	\textbf{62.7}&	\textbf{92.6}&	\textbf{69.3}&	\textbf{53.4}&	\textbf{42.9}&	\textbf{37.6}&	\textbf{59.2}\\
\cline{1-13}

\multirow{1}{*}{Asm2vec$^{\ast}$} 	&74.1&	29.7&	18.1&	12.1&	9.1&	28.6&	79.5&	41.8&	31.2&	22.7&	17.9&	38.6\\
\multirow{1}{*}{Asm2vec$^{\ast}$+\binenhance{}} 	&\textbf{94.9}&	\textbf{67.0}&	\textbf{51.8}&	\textbf{41.7}&	\textbf{36.5}&	\textbf{58.4}&	\textbf{96.6}&	\textbf{74.6}&	\textbf{55.9}&	\textbf{42.5}&	\textbf{36.9}&	\textbf{61.3}\\
\cline{1-13}

\multirow{1}{*}{Asteria} &94.5&	77.2&	60.9&	48.2&	40.1&	64.2&	97.1&	82.1&	63.8&	47.3&	36.9&	65.4\\
\multirow{1}{*}{Asteria+\binenhance{}} &\textbf{95.8}&	\textbf{82.4}&	\textbf{72.3}&	\textbf{62.3}&	\textbf{55.1}&	\textbf{73.6}&	\textbf{98.3}&	\textbf{86.1}&	\textbf{72.9}&	\textbf{58.8}&	\textbf{49.4}&	\textbf{73.1} \\
\cline{1-13}

\multirow{1}{*}{HermesSim} &98.9&	94.1&	84.8&	73.1&	64.9&	83.2&	97.5&	89.8&	74.7&	56.8&	46.2&	73.0 \\
\multirow{1}{*}{HermesSim+\binenhance{}}  &\textbf{98.9}&	\textbf{94.9}&	\textbf{87.4}&	\textbf{77.5}&	\textbf{70.1}&	\textbf{85.8}&	\textbf{98.7}&	\textbf{92.1}&	\textbf{79.8}&	\textbf{63.4}&	\textbf{53.2}&	\textbf{77.4} \\
\bottomrule[2pt]
\end{tabular}
\begin{tablenotes}
\item {\small $\ast$ Asm2vec only support X86 architecture.}
\end{tablenotes}
\end{threeparttable}
\label{table:map_eval}
\end{table*}

In real-world binary code search tasks, the size of the function pool is often significantly larger than the pool size of the experiments.
To evaluate the impact of increasing pool size, we set the function pool sizes to (2, 16, 32, 128, 512, 1024, 2048, 4096, 8192, 10000).
Table \ref{table:map_eval} illustrates the MAP scores of five baseline methods and their enhancement via \binenhance{} across two distinct public datasets ($D1$, $D2\_norm$). 
As shown in Table~\ref{table:map_eval}, the enhancement versions consistently outperform the baselines in all the function pool sizes, with maximal MAP improvements of 27.4\% and 28.3\% on the respective datasets, and an average MAP improvement of 16.1\% (from 53.6\% to 69.7\%). 
Notably, some methods after using \binenhance{} even surpassed originally more advanced methods. For instance, in $D1$ with the function pool size of 10000, Asm2vec's MAP score increases from 9.1\% to 36.5\%, overtaking TREX's 21.4\%.
It indicates that while the existing methods inherently possess the capability to capture some complex semantic information, they lack an efficient mechanism for integrating and utilizing this information. 
\binenhance{} addresses this gap by systematically curating and amalgamating that semantic information through EESG.

\begin{figure}
    \centering
    \includegraphics[width=\linewidth, trim=0cm 0cm 0cm 1.6cm, clip]{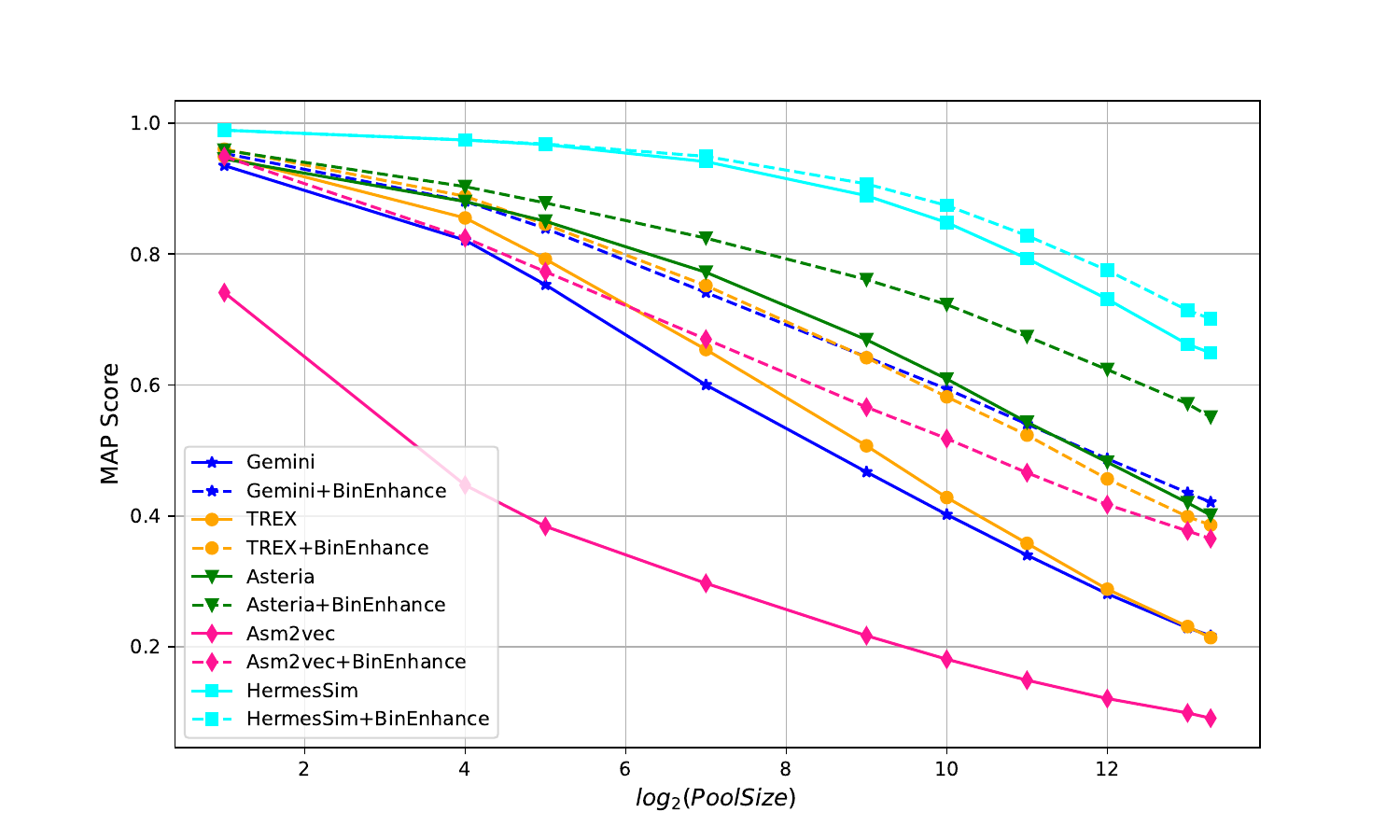}
    \vspace{-1.3em}
    \caption{MAP scores in different pool sizes ($D1$) }
    \label{fig:p1}
\end{figure}

\begin{figure}
    \centering
    \includegraphics[width=\linewidth, trim=0cm 0cm 0cm 1.6cm, clip]{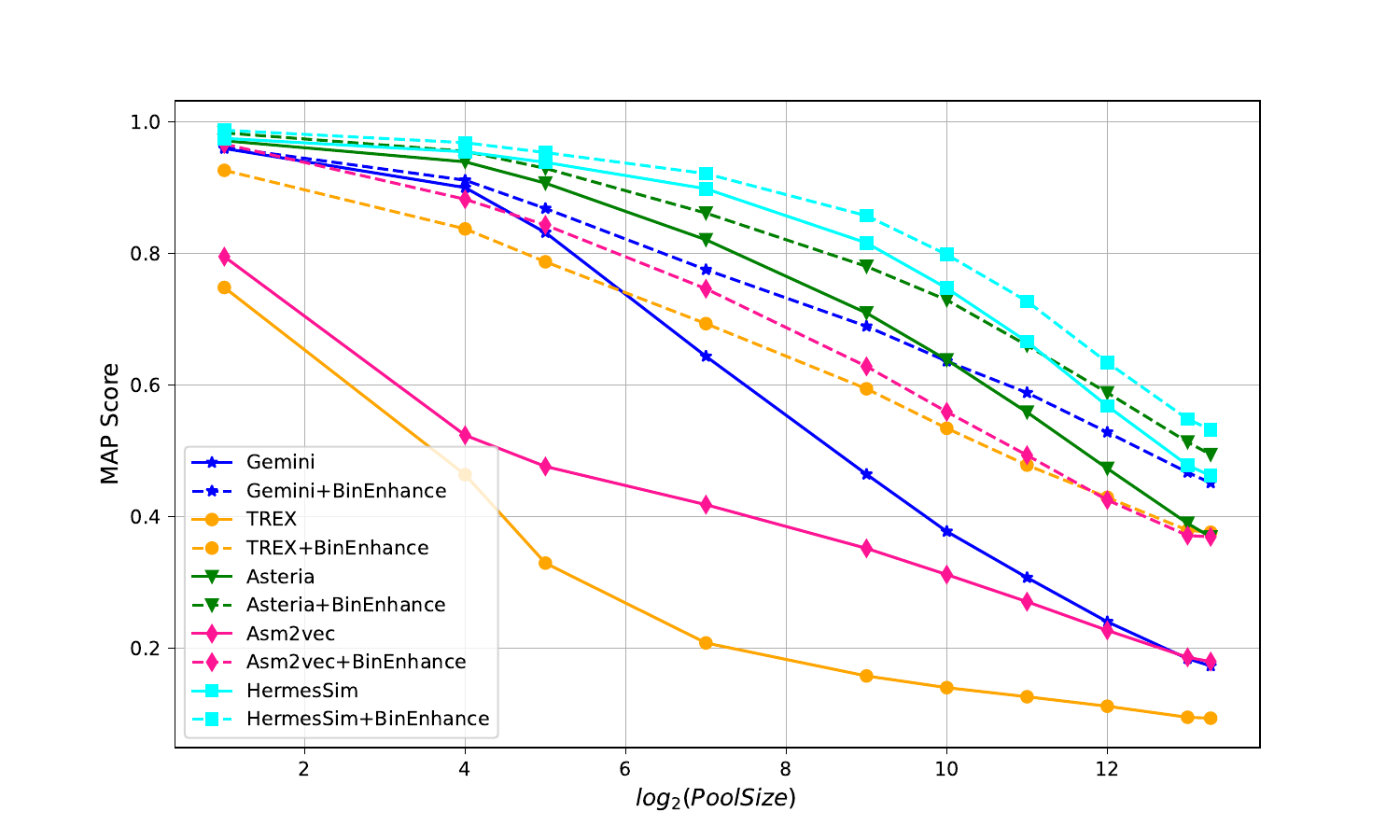}
    \vspace{-1.3em}
    \caption{MAP scores in different pool sizes ($D2\_norm$) }
    \label{fig:p2}
\end{figure}

As illustrated in Figure~\ref{fig:p1} and~\ref{fig:p2}, there is a notable decline in the MAP scores across all baselines as the size of the function pool expands.
This trend highlights the increasing challenge of discerning and recalling homologous functions amidst a growing array of options. 
However, it's particularly noteworthy that the disparity in MAP scores between the \binenhance{} model (represented by a dashed line) and the original model (represented by a solid line) progressively widens. 
For instance, within a function pool of just two functions, the original Asteria model scores 94.5\% in $D1$, which increases to 95.8\% after \binenhance{} enhancement (an increase of 1.3\%). This gap further escalates to 15\% when the function pool increases to 10,000. 
This widening margin underscores \binenhance{}'s superior stability and accuracy in recalling homologous functions from a large function pool, making it more practical. 
Thanks to the implementation of function embedding enhancement and similarity combination, \binenhance{} effectively captures and leverages the subtle semantic similarities between homologous functions, thereby achieving markedly improved performance.


\begin{tcolorbox}
\vspace{-5pt}
\textbf{Answer to RQ.1.} \binenhance{} demonstrates significant improvement across all the baselines, evidenced by the average increment in MAP scores on the two public datasets (16.1\%, from 53.6\% to 69.7\%). Furthermore, its improvement for each baseline is positively correlated with the size of the function pool.
\vspace{-5pt}
\end{tcolorbox}

\subsection{RQ2: Robustness Evaluation}

\begin{table*}[t]
\centering
\vspace{-0.5em}
\caption{MAP scores of different methods in different optimization options and architectures ($D1$).}
\vspace{-0.7em}
\renewcommand{\arraystretch}{1.1}
\setlength {\tabcolsep} {4pt}

\begin{threeparttable}
\begin{tabular}{l|ccc|cccccc}
\toprule[2pt]
\multicolumn{1}{c|}{\multirow{2}{*}{\textbf{Models}}} & 
\multicolumn{3}{c|}{\textbf{Cross-architecture}}&\multicolumn{6}{c}{\textbf{Cross-optimization option}}\\
\cline{2-4}
\cline{4-10}
\multicolumn{1}{c|}{} & \textbf{ARM,X86} & \textbf{ARM,X64} & \textbf{X64,X86} & \textbf{O0,O1} & \textbf{O0,O2} & \textbf{O0,O3} & \textbf{O1,O2} & \textbf{O1,O3} & \textbf{O2,O3}\\
\hline


\multirow{1}{*}{Gemini}  &33.6&	32.9&	52.2&	35.5&	30.6&	25.2&	53.3&	42.0&	54.0
\\
\multirow{1}{*}{Gemini+\binenhance{}}  &\textbf{58.2}&	\textbf{63.0}&	\textbf{76.0}&	\textbf{58.3}&	\textbf{51.4}&	\textbf{42.1}&	\textbf{68.6}&	\textbf{58.5}&	\textbf{68.0}
\\
\cline{1-10}

\multirow{1}{*}{TREX}  &29.4&	32.4&	56.5&	42.1&	37.0&	29.2&	71.1&	58.1&	69.0

\\
\multirow{1}{*}{TREX+\binenhance{}}  &\textbf{66.3}&	\textbf{68.3}&	\textbf{83.9}&	\textbf{67.1}&	\textbf{61.0}&	\textbf{55.2}&	\textbf{84.2}&	\textbf{74.3}&	\textbf{74.8} 

\\
\cline{1-10}

\multirow{1}{*}{Asm2vec$^{\ast}$} &-	&-	&-	&	1.5&	1.6&	1.7&	29.4&	25.7&	43.4

\\
\multirow{1}{*}{Asm2vec$^{\ast}$+\binenhance{}}  &-	&-	&-	&	\textbf{43.4}&	\textbf{42.3}&	\textbf{35.1}&	\textbf{68.4}&	\textbf{52.3}&	\textbf{64.6}
 
\\
\cline{1-10}
\multirow{1}{*}{Asteria} &68.2&	60.2&	69.3&	58.9&	49.4&	42.0&	57.4&	45.9&	57.7

\\
\multirow{1}{*}{Asteria+\binenhance{}}  &\textbf{79.4}&	\textbf{79.4}&	\textbf{81.2}&	\textbf{73.2}&	\textbf{61.2}&	\textbf{56.5}&	\textbf{77.9}&	\textbf{63.4}&	\textbf{71.6} 
\\
\cline{1-10}

\multirow{1}{*}{HermesSim}  &90.7&	89.9&	89.9&	83.1&	84.7&	79.5&	81.9&	82.4&	79.2
 
\\
\multirow{1}{*}{HermesSim+\binenhance{}}  &\textbf{96.3}&	\textbf{96.2}&	\textbf{97.4}&	\textbf{91.5}&	\textbf{91.6}&	\textbf{87.1}&	\textbf{96.2}&	\textbf{91.9}&	\textbf{91.6}
\\



\bottomrule[2pt]
\end{tabular}

\begin{tablenotes}
\item {\small $\ast$ Asm2vec do not support cross-architecture.}
\end{tablenotes}

\end{threeparttable}
\vspace{-1.0em}
\label{table:c1}
\end{table*}

\begin{table*}[t]
\centering
\vspace{-0.5em}
\caption{MAP scores of different methods in different optimization options and architectures ($D2\_norm$).}
\vspace{-0.7em}
\renewcommand{\arraystretch}{1.1}
\setlength {\tabcolsep} {4pt}

\begin{threeparttable}
\begin{tabular}{l|ccc|cccccc}
\toprule[2pt]
\multicolumn{1}{c|}{\multirow{2}{*}{\textbf{Models}}} & 
\multicolumn{3}{c|}{\textbf{Cross-architecture}}&\multicolumn{6}{c}{\textbf{Cross-optimization option}}\\
\cline{2-4}
\cline{4-10}
\multicolumn{1}{c|}{} & \textbf{ARM,X86} & \textbf{ARM,MIPS} & \textbf{MIPS,X86} & \textbf{O0,O1} & \textbf{O0,O2} & \textbf{O0,O3} & \textbf{O1,O2} & \textbf{O1,O3} & \textbf{O2,O3}\\
\hline


\multirow{1}{*}{Gemini}  &23.6&	21.9&	32.0&		26.4&	22.7&	18.9&	54.2&	45.8&	59.1
\\
\multirow{1}{*}{Gemini+\binenhance{}}  &\textbf{58.5}&	\textbf{60.8}&	\textbf{58.6}&		\textbf{55.3}&	\textbf{54.6}&	\textbf{51.9}&	\textbf{83.8}&	\textbf{76.0}&	\textbf{81.4}
\\
\cline{1-10}

\multirow{1}{*}{TREX}  &3.1&	1.7&	1.5&		15.3&	13.3&	12.7&	65.5&	60.5&	76.1

\\
\multirow{1}{*}{TREX+\binenhance{}}  &\textbf{56.1}&	\textbf{26.1}&	\textbf{36.2}&		\textbf{55.3}&	\textbf{52.0}&	\textbf{48.9}&	\textbf{84.4}&	\textbf{76.1}&	\textbf{84.9}

\\
\cline{1-10}

\multirow{1}{*}{Asm2vec$^{\ast}$} &-	&-	&-	&		4.4&	3.7&	3.5&	52.7&	47.7&	70.2

\\
\multirow{1}{*}{Asm2vec$^{\ast}$+\binenhance{}}  &-	&-	&-	&		\textbf{46.1}&	\textbf{44.2}&	\textbf{42.2}&	\textbf{73.3}&	\textbf{68.1}&	\textbf{79.4}
 
\\
\cline{1-10}
\multirow{1}{*}{Asteria} &57.3&	55.5&	52.1&		40.0&	33.7&	29.6&	60.8&	55.8&	70.6

\\
\multirow{1}{*}{Asteria+\binenhance{}}  &\textbf{66.0}&	\textbf{69.2}&	\textbf{67.0}&		\textbf{56.3}&	\textbf{55.1}&	\textbf{52.6}&	\textbf{80.0}&	\textbf{75.6}&	\textbf{87.3}
\\
\cline{1-10}

\multirow{1}{*}{HermesSim}  &60.3&	63.4&	79.1&		76.5&	74.4&	73.1&	82.1&	82.5&	83.9
 
\\
\multirow{1}{*}{HermesSim+\binenhance{}}  &\textbf{74.3}&	\textbf{75.9}&	\textbf{87.5}&		\textbf{88.2}&	\textbf{85.2}&	\textbf{82.6}&	\textbf{94.4}&	\textbf{92.2}&	\textbf{95.6}
\\



\bottomrule[2pt]
\end{tabular}

\begin{tablenotes}
\item {\small $\ast$ Asm2vec do not support cross-architecture.}
\end{tablenotes}

\end{threeparttable}
\vspace{-1.0em}
\label{table:c2}
\end{table*}

In this section, we conduct extensive experiments to asses \binenhance{}'s ability to improve baseline models under various architectures and compilation optimization options, aiming to verify its adaptability in real-world scenarios. 
Cross-architecture experiments are conducted on the $O2$ level compilation optimization option, and cross-optimization option experiments are conducted on $X86$ architecture. 
The results, detailed in Tables~\ref{table:c1} and~\ref{table:c2}, show \binenhance{}'s significant enhancements in cross-architecture and cross-optimization option experiments. 
Specifically, \binenhance{}, when applied in HermesSim, has the best results in cross-architecture and cross-optimization option tasks in all the public datasets.
However, using the Transformer model without a specific cross-architecture design, TREX lagged behind Gemini in these tasks but mostly showed superior performance in cross-compilation optimizations, especially with \binenhance{}.

\begin{figure}
    \centering
    \vspace{-0.8em}
    \includegraphics[width=\linewidth, trim=0cm 0cm 0cm 1cm, clip]{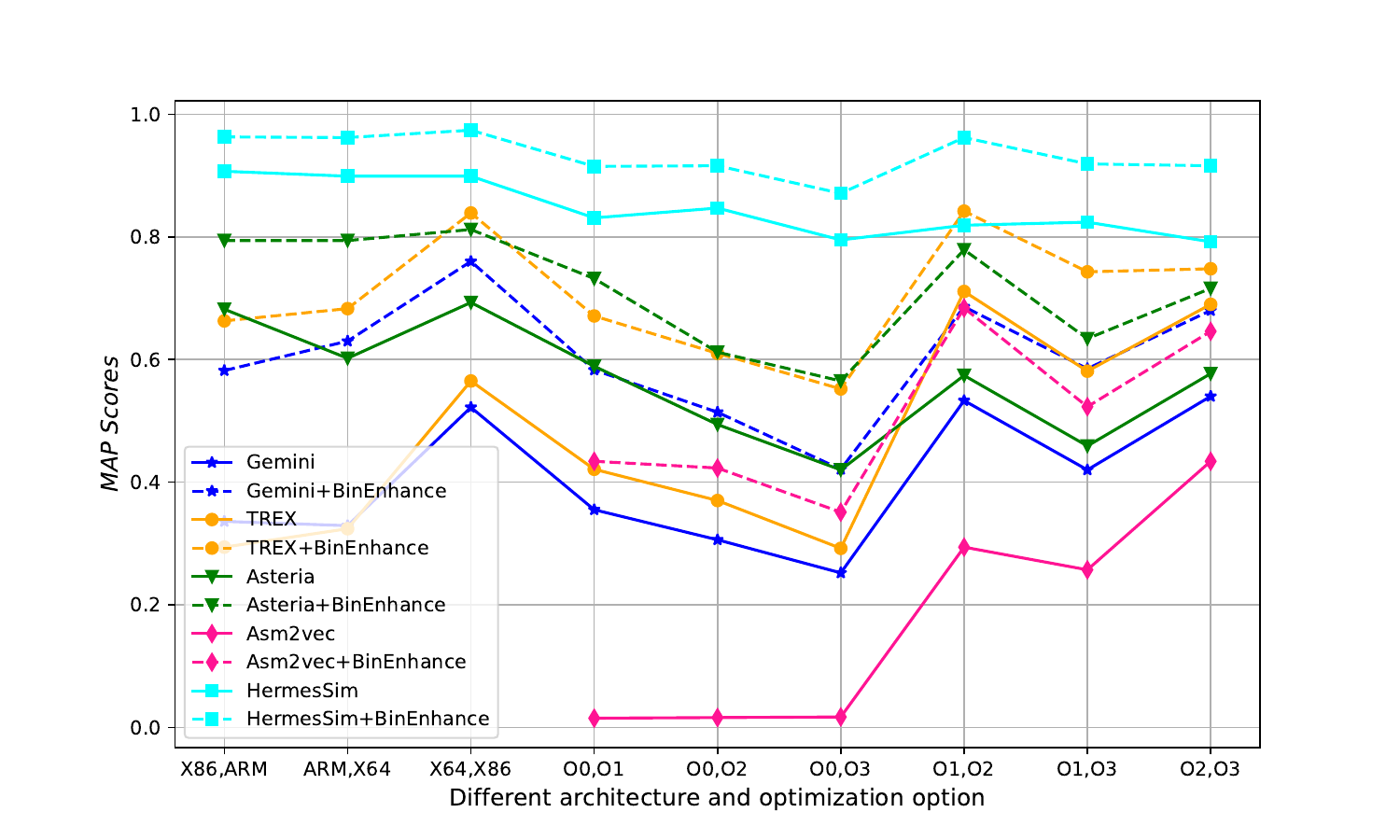}
    \vspace{-2.3em}
    \caption{MAP scores in different compilation settings ($D1$) }
    \vspace{-2.0em}
    \label{fig:c1}
\end{figure}

\begin{figure}
    \centering
    \vspace{-0.8em}
    \includegraphics[width=\linewidth, trim=0cm 0cm 0cm 1cm, clip]{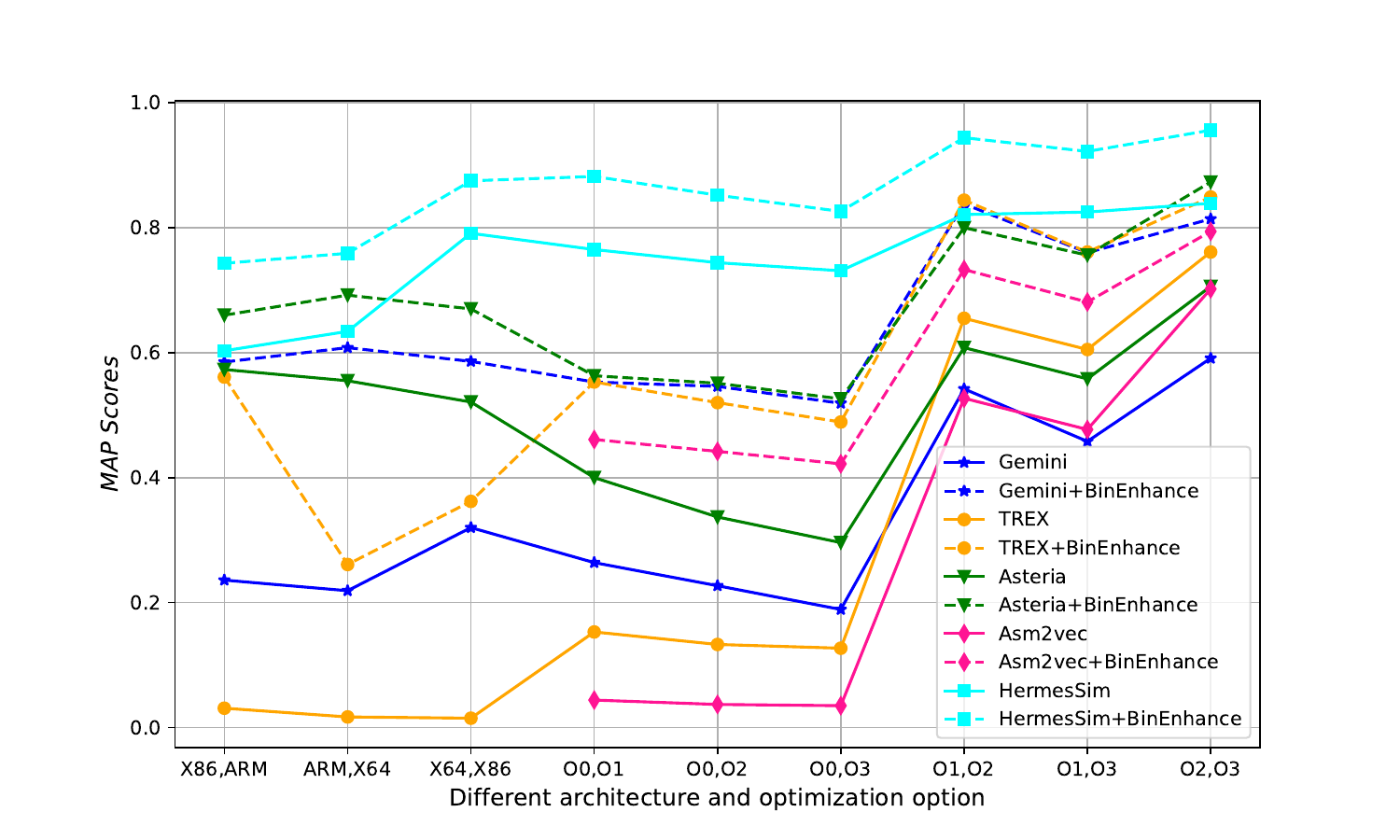}
    \vspace{-2.3em}
    \caption{MAP in different compilation settings ($D2\_norm$) }
    \vspace{-2.0em}
    \label{fig:c2}
\end{figure}

Figure~\ref{fig:c1} and~\ref{fig:c2} show that the improvement of the baseline models caused by BinEnhence is mostly stable across architectures and compilation optimization options (the interval between dashed and solid lines mostly does not vary significantly). 
In some tasks with MAP improvement relatively larger changes, there is no optimization strategy for the function in the $O0$ case, while there are some optimization strategies that change the structure such as CFG under the other optimization options, resulting in a very low score for the Asm2vec original model in the tasks (functions under $O0$ case recall functions under other options). 
\binenhance{} mitigates the effects of the optimization strategy, and the boosts are larger in the $O0$ case than other optimization options, so there are some fluctuations of Asm2vec.
These results underscore \binenhance{}'s robustness and consistency in diverse compilation environments.
Its stability stems from the EESG's ability to create a consistent external environment for homologous functions, which combined with the SEM model, ensures a reliable performance enhancement across all test tasks.

\begin{tcolorbox}
\vspace{-5pt}
\textbf{Answer to RQ.2.} \binenhance{} stably enhances baselines across cross-architecture and cross-optimization option tasks, without succumbing to significant performance dips under various compilation settings.
\vspace{-5pt}
\end{tcolorbox}

\subsection{RQ3: Impact of function inlining}

\begin{table*}[t]
\centering
\vspace{-0.5em}
\renewcommand{\arraystretch}{1.1}
\setlength {\tabcolsep} {2.5pt}
\caption{MAP scores of different methods of binary code search tasks in inlining and non-inlining compilation settings.}
\vspace{-0.7em}
\begin{threeparttable}
\begin{tabular}{l|cccccc|cccccc}
\toprule[2pt]
\multicolumn{1}{c|}{\multirow{2}{*}{\textbf{Models}}} & 
\multicolumn{6}{c|}{\textbf{Normal-to-Normal}} & 
\multicolumn{6}{c}{\textbf{Normal-to-Noinline}} \\

\cline{2-13}

\multicolumn{1}{c|}{} & \textbf{P=2} & \textbf{P=128} & \textbf{P=1024}&\textbf{P=4096} & \textbf{P=10000} & \textbf{Avg.} & \textbf{P=2} & \textbf{P=128} & \textbf{P=1024}&\textbf{P=4096} & \textbf{P=10000} & \textbf{Avg.}\\
\hline


\multirow{1}{*}{Gemini} &90.6&	57.5&	38.0&	27.6&	21.7&	47.1&	89.6&	50.3&	28.4&	18.9&	14.0&	40.2\\
\multirow{1}{*}{Gemini+\binenhance{}} &\textbf{93.6}&	\textbf{73.6}&	\textbf{62.7}&	\textbf{56.1}&	\textbf{51.3}&	\textbf{67.5}&	\textbf{92.3}&	\textbf{69.6}&	\textbf{57.1}&	\textbf{49.0}&	\textbf{44.9}&	\textbf{62.6}\\
\cline{1-13}

\multirow{1}{*}{TREX} &73.0&	24.3&	16.3&	13.4&	11.9&	27.8&	71.9&	22.1&	12.7&	9.6&	8.1&	24.9\\
\multirow{1}{*}{TREX+\binenhance{}} &\textbf{92.8}&	\textbf{70.4}&	\textbf{60.4}&	\textbf{53.0}&	\textbf{48.7}&	\textbf{65.1}&	\textbf{91.0}&	\textbf{66.7}&	\textbf{52.8}&	\textbf{44.0}&	\textbf{41.4}&	\textbf{59.2}\\
\cline{1-13}

\multirow{1}{*}{Asm2vec$^{\ast}$} 	&86.7&	40.7&	30.2&	23.8&	19.3&	40.1&	11.2&	6.0&	3.0&	2.3&	1.7&	4.8\\
\multirow{1}{*}{Asm2vec$^{\ast}$+\binenhance{}} 	&\textbf{95.0}&	\textbf{72.9}&	\textbf{60.3}&	\textbf{50.5}&	\textbf{46.6}&	\textbf{65.1}&	\textbf{92.9}&	\textbf{62.6}&	\textbf{47.7}&	\textbf{39.1}&	\textbf{37.0}&	\textbf{55.9}\\
\cline{1-13}

\multirow{1}{*}{Asteria} &94.9&	71.9&	56.8&	47.2&	39.2&	62.0&	54.7&	29.2&	21.3&	17.5&	14.1&	27.4\\
\multirow{1}{*}{Asteria+\binenhance{}} &\textbf{97.1}&	\textbf{77.7}&	\textbf{67.8}&	\textbf{59.2}&	\textbf{53.4}&	\textbf{71.0}&	\textbf{59.1}&	\textbf{49.2}&	\textbf{38.2}&	\textbf{31.3}&	\textbf{27.1}&	\textbf{41.0} \\
\cline{1-13}

\multirow{1}{*}{HermesSim} &93.9&	85.3&	74.5&	62.7&	53.3&	73.9&	92.7&	84.5&	72.0&	57.6&	48.6&	71.1 \\
\multirow{1}{*}{HermesSim+\binenhance{}}  &\textbf{97.9}&	\textbf{87.9}&	\textbf{79.1}&	\textbf{69.6}&	\textbf{61.7}&	\textbf{79.2}&	\textbf{97.6}&	\textbf{87.1}&	\textbf{75.9}&	\textbf{64.5}&	\textbf{56.4}&	\textbf{76.3} \\
\bottomrule[2pt]
\end{tabular}
\begin{tablenotes}
\item {\small $\ast$ Asm2vec only support X86 architecture.}
\end{tablenotes}
\vspace{-2.0em}
\end{threeparttable}
\label{table:inline}
\end{table*}
\begin{figure}
    \centering
    \includegraphics[width=\linewidth]{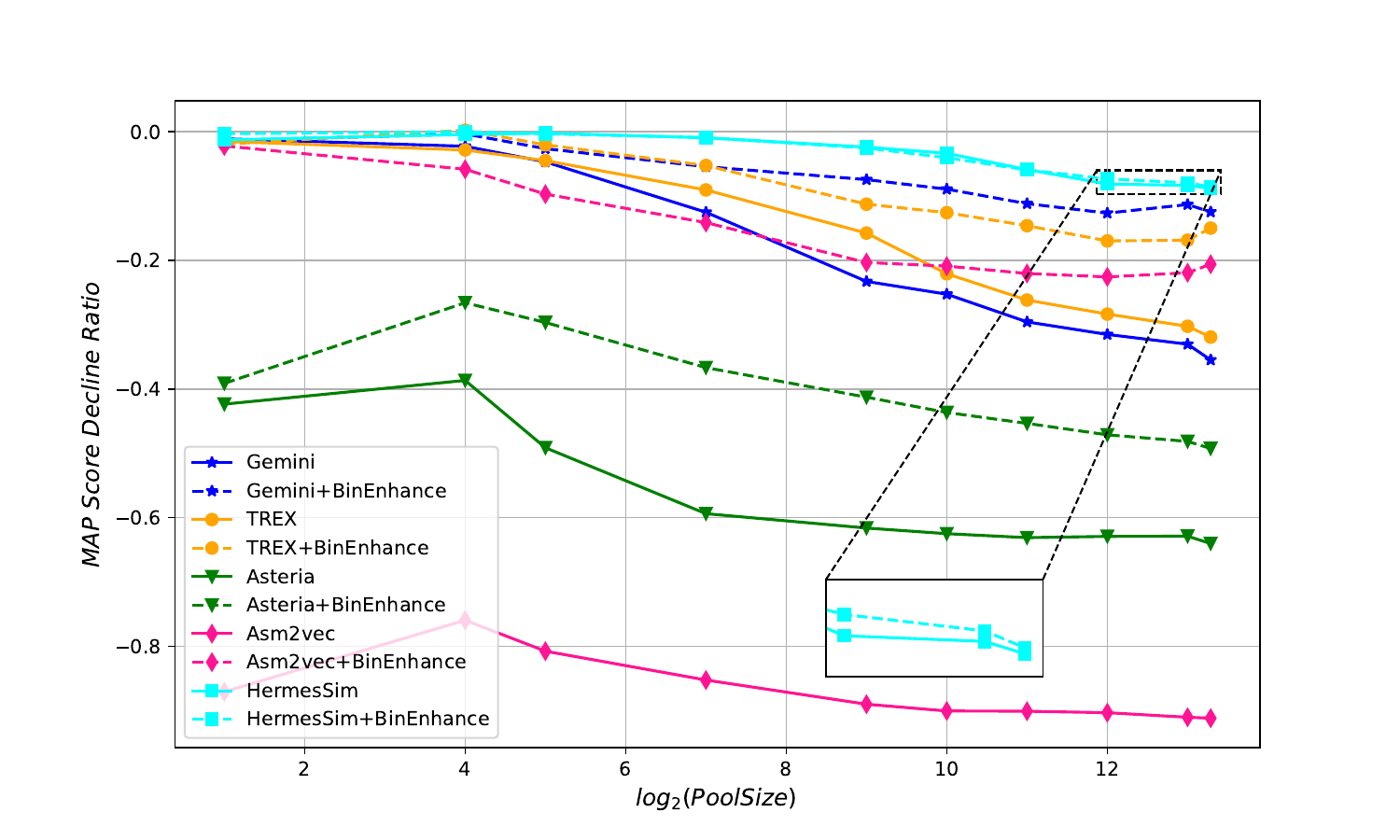}
    \vspace{-2.3em}
    \caption{MAP Score Decline Ratio in function inlining }
    \vspace{-1.5em}
    \label{fig:inline}
\end{figure}
\begin{figure}
    \centering
    \vspace{-1.0em}
    \includegraphics[width=\linewidth]{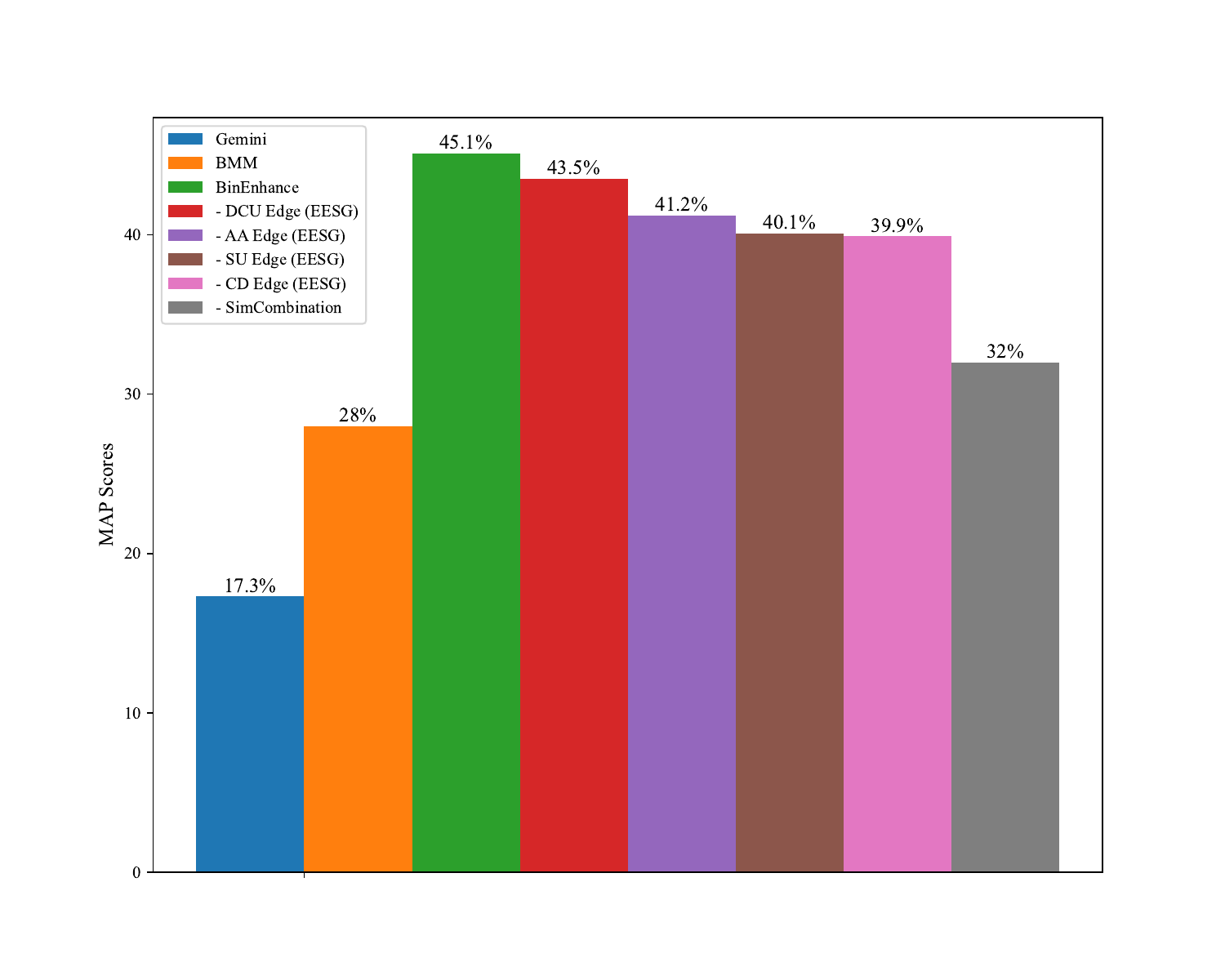}
     \vspace{-3.8em}
    \caption{MAP scores of different situations}
    \label{fig:as}
    \vspace{-2.5em}
\end{figure}
In this section, we explore the effects of \binenhance{} on function inlining, employing datasets $D2\_norm$ and $D2\_noinline$ for this purpose.
To evaluate the impact of function inlining, we first extract homologous functions with and without inlining from the two datasets. 
This extraction is accomplished by analyzing the function call graph and the corresponding symbol information provided by the datasets.
Based on these lists of homologous functions, we reconstruct two types of homologous function pairs: \textbf{Normal-to-Normal} ($NI$, both functions are from the normally compiled datasets $D2\_norm$) and \textbf{Normal-to-Noinline} ($NO$, one function from the $D2\_norm$ and the other from the without function inlining datasets $D2\_noinline$). 
The functions in both homologous function pairs are identical. 
We evaluate these two types of homologous function pairs separately, as shown in Table~\ref{table:inline}, which presents the results for two different tasks.

Firstly, whether in $NI$ or $NO$ tasks, \binenhance{} significantly improves MAP scores compared to the baseline method.
Since the only difference between the two homologous function pairs is the presence or absence of inlining, the impact of function inlining can be evaluated by the MAP difference between $NI$ and $NO$.
We use the MAP score decline ratio = $(MAP_{NI}-MAP_{NO})/MAP_{NI}$ as the measurement standard.
For clarity, Figure~\ref{fig:inline} illustrates the effect of \binenhance{} on function inlining, with the y-axis representing the MAP score decline ratio, which indicates the MAP score loss introduced by function inlining.
We observed that, after applying \binenhance{}, most baseline methods showed a reduced performance loss (the dashed line is located above the solid line).
Both the original model of HermesSim and the model enhanced by \binenhance{} exhibited a relatively small decrease in MAP when dealing with function inlining. 
HermesSim differs from most other methods that focus on binary code design. 
It uses a compiler-independent toy IR~\cite{Ghidra} to convert it into SOG and then employs multi-head softmax to selectively learn the semantics of different nodes on SOG, which is constrained by the analysis tool~\cite{Ghidra}.
Conversely, Asm2vec is significantly affected by function inlining, but \binenhance{} greatly mitigates its MAP score loss. These results are attributed to our proposed stable EESG for cross-compilation settings.

\begin{tcolorbox}
\vspace{-5pt}
\textbf{Answer to RQ.3.} Function inlining leads to a substantial performance loss in the binary code search task, but \binenhance{} mitigates the decline and improves the baseline model's ability to cope with optimization strategies such as function inlining. 
\vspace{-5pt}
\end{tcolorbox}

\subsection{RQ4: Ablation Study}\label{ablation}

To evaluate the impact of different types of edge in EESG and similarity combination on the performance of the \binenhance{} framework, we design different ablation experiments on the test dataset separately. For clarity of comparison and time cost, we select Gemini as an example, Figure~\ref{fig:as} shows the results of all experiments including:

\begin{itemize}[leftmargin=10pt]
\item \textbf{Gemini}: the original model has a MAP score of 17.3\%.
\item \textbf{\binenhance{}}: indicates the complete framework of \binenhance{} with a MAP score of 45.1\%.
\item \textbf{-$CD$ Edge}: EESG does not exist $CD$ edges, with a reduced MAP score of 5.2\%, which supplements the semantics through function calls, mitigating the performance loss.
\item \textbf{-$DCU$ Edge}: EESG does not exist $DCU$ edges, with a reduced MAP score of 1.5\%. Global data does not exist in all functions, so its contribution is relatively smaller.
\item \textbf{-$AA$ Edge}: EESG does not exist $AA$ edges, with a reduced MAP score of 3.9\%, which establishes edges between multiplexed functions through location information, supplementing the semantics with location semantic information.
\item \textbf{-$SU$ Edge}: EESG does not exist $SU$ edges, with a reduced MAP score of 5.0\%, which demonstrates that our SEM can merge string semantics with function semantics.
\item \textbf{-SimCombination}: \binenhance{} does not use the similarity combination module, with a reduced MAP score of 13.1\%, it decreases the most, which illustrates that it is effective for the task when data features exist.
\end{itemize}
In addition, we select BMM \cite{guo2022exploring} for evaluating the performance of function call relationship methods, and the final score is 28.0\%, which is 17.1\% lower than our method. This also proves that our EESG is superior to function call graphs.

\begin{tcolorbox}
\vspace{-5pt}
\textbf{Answer to RQ.4.} Each component of \binenhance{} plays a crucial role in the final result, and the absence of any one of them can lead to a degradation in the performance of the binary search task.
\vspace{-5pt}
\end{tcolorbox}

\subsection{RQ5: Efficiency Evaluation}\label{rq5}
\begin{table}
  \setlength {\tabcolsep} {2pt}
  \renewcommand{\arraystretch}{1.3}
  \caption{Efficiency evaluation}
  \label{tab:ast_time}
  
  \begin{tabular}{l|ccccc}
    \toprule[2pt]
    Model & Feature & Train & Embed &  Search & All  \\
    \cline{1-6}
    Asteria & 2.14 &86.40  &  1.21 &  149.20 & 238.95\\
    \cline{1-6}
    \binenhance{} & \textbf{1.22}  &\textbf{2.25} & \textbf{0.84} & \textbf{118.32} & \textbf{122.63}\\
    \cline{1-6}
    Asteria+\binenhance{}  &2.14 & 88.65  & 2.05 & \textbf{118.32} & 211.16\\
  \bottomrule[2pt]
\end{tabular}
\vspace{-1.11em}
\end{table}

\begin{figure}
    \centering
    \includegraphics[width=\linewidth]{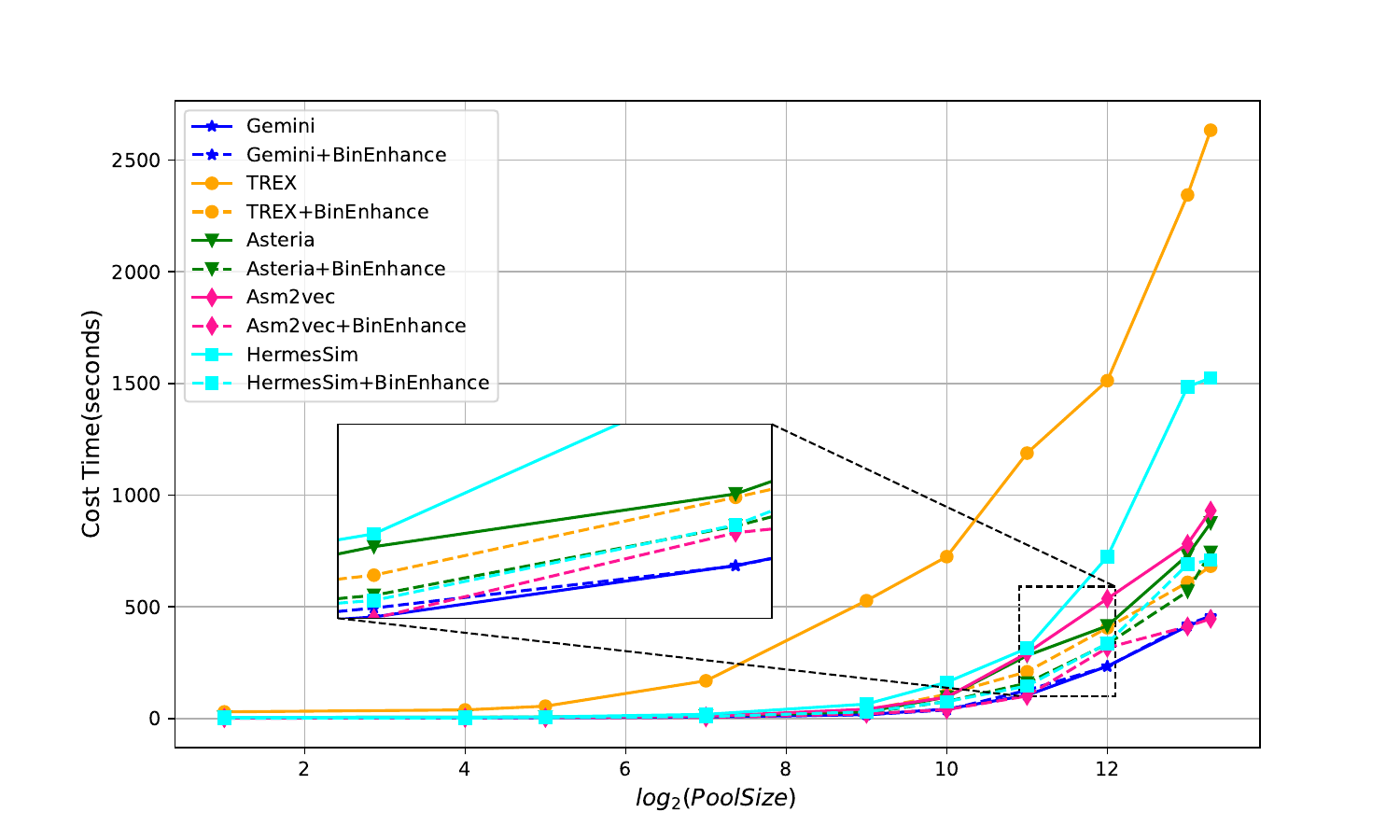}
    \caption{The impact of embedding dimension on time cost.}
    \label{fig:ct}
\end{figure}

Table~\ref{tab:ast_time} shows the time cost required for each step of processing a binary file for Asteria as well as \binenhance{}.
The additional time cost incurred by \binenhance{} is mainly in the training (2.25s) and the generating function embedding (0.84s), which is because these two operations cannot be performed in parallel with the baseline method (feature extraction can be operated in parallel), but in reality, the extra time cost is not high compared to the baseline method.
In addition, \binenhance{} ends up producing smaller function embedding dimensions compared to the baseline method. 
As a result, the time cost spent on binary code search by our method is drastically reduced, as shown in Figure~\ref{fig:ct}. 
It shows that as the function pool size increases, the time spent on binary code search increases immediately. 
It is worth mentioning that, as shown in Table~\ref{tab:dim}, for all the baseline methods, we have reduced the function embedding dimension for almost all of them. More specifically, the function embeddings of \binenhance{}, when applied to the binary code search task, take less time (even less than 1/4) compared to the original methods, and the saved time increases as the function pool size increases.

\begin{tcolorbox}
\vspace{-5pt}
\textbf{Answer to RQ.5.} The additional time cost of \binenhance{} is in training and generating function embeddings, but it significantly reduces the time for binary code search tasks, resulting in an overall lower time cost compared to the original model.
\vspace{-5pt}
\end{tcolorbox}

\subsection{RQ6: 1-day Vulnerability Search}
\begin{table}
  \centering
  \caption{Results of vulnerability search (\textit{fail to recall}).}
  \label{tab:real}
  \scriptsize
  \begin{tabularx}{\columnwidth}{p{0.001\columnwidth}cccccccc}
    \toprule
    \midrule
      ID & CVE & Tot. & Gemini & Asteria & TREX & HermesSim & Ours\\
    \midrule
    1 & 2014-4877&20 &3 &19 &14 &0 &0 \\
    2 & 2016-8622 & 9& 9&8 &7 &8 &2 \\
    3 & 2016-6301 &10 &10 &0 &10 &0 &0  \\
    4 & 2016-8618&12 &11 &11 &6 &6 &3  \\
    5 & 2018-19519&6 &6 &6 &6 &0 &0  \\
    6 & 2018-1000301 &1 &0 &0 &0 &0 &0  \\
    7 & 2018-16230&{7}&{4}&{4}&{6}&{3}&{3} \\
    8 & {2018-16452}&{20}&{1}&{19}&{19}&{2}&{1} \\
    9 & {2018-16451}&{3}&{3}&{2}&{3}&{2}&{2} \\
    10 & {2020-8306}&{20}&{1}&{18}&{18}&{4}&{2} \\
    11 & 2021-22924&3 &2 &3 &1 &2 &2 \\
    12 & {2022-0778}&{5}&{5}&{5}&{5}&{0}&{0}\\
     \midrule
     &MAP &- &14.2 &26.8 &15.2 &60.2 &\textbf{67.9}   \\
\midrule
  \bottomrule
\end{tabularx}
\vspace{-1.6em}
\end{table}

\begin{figure}
    \centering
    \includegraphics[width=\linewidth]{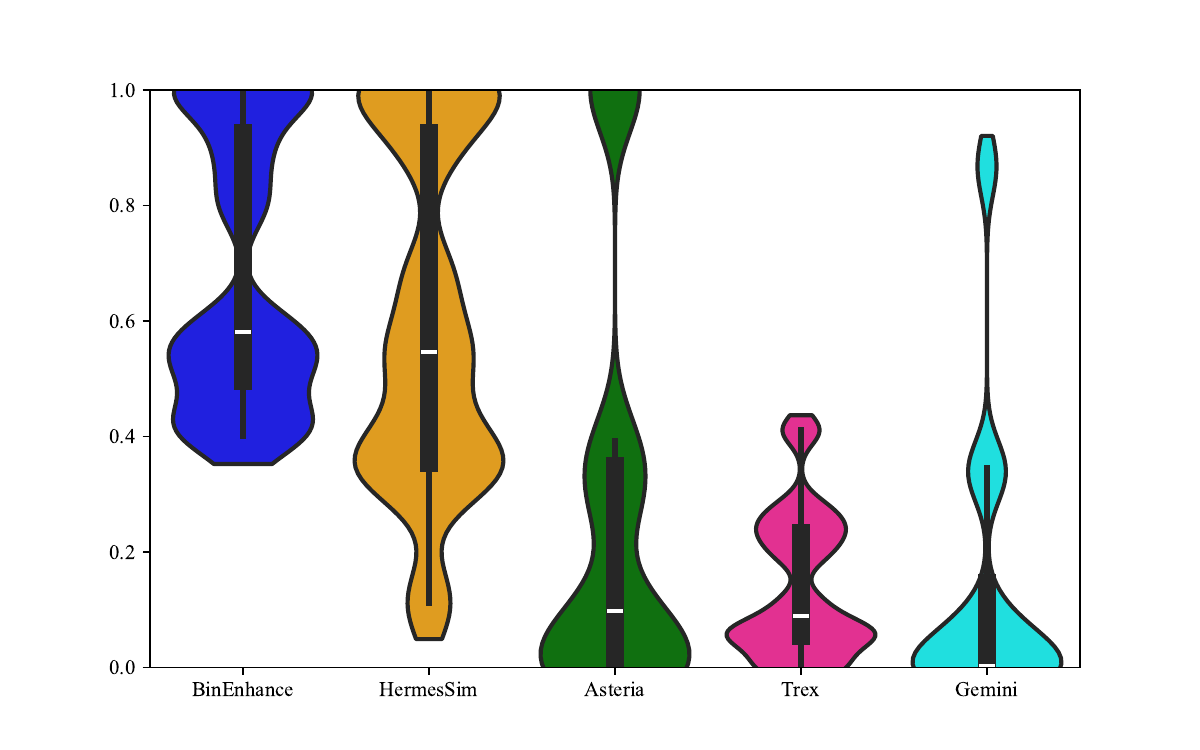}
    \vspace{-1.8em}
    \caption{The violin plot for each method on vulnerability search.}
    \vspace{-1.5em}
    \label{fig:real}
\end{figure}

In this section, we utilized the $D3\_firmware$ alongside CVE vulnerability functions dataset to determine the effectiveness of the \binenhance{} in enhancing real-world 1-day vulnerability search capabilities.
To ensure comprehensive coverage of both sources~\cite{luovulhawk, yang2021asteria}, we selected 12 high-impact and commonly reported CVE vulnerabilities~\cite{Cve-cve-2018-1000301, Cve-cve-2016-8618, Cve-cve-2016-8622, Cve-cve-2014-4877, Cve-cve-2021-22924, Cve-cve-2018-19519, Cve-cve-2016-6301, Cve-cve-2020-8036, Cve-cve-2022-0778, Cve-cve-2018-16230, Cve-cve-2018-16451, Cve-cve-2018-16452}, focusing on widely used open-source components (e.g., $OpenSSL$, $Libcurl$).
The evaluation involves homologous function recall using HermesSim and \binenhance{} for each CVE vulnerability function and those undetected in the decompressed firmware binaries.
Like the HermesSim experiment, we identify the ground truth by manually
examining the top-20 results of all methods and then we obtain 116 vulnerability functions from $D3\_firmware$.
As shown in Table~\ref{tab:real}, \binenhance{} recalls 101 vulnerability functions with a MAP score of 67.9\%, which is 12 more than the recall number of HermesSim with an increased MAP score of 7.7\%. \binenhance{} Through incorporating external semantic information that contrasts with the similar internal code structure of these non-homologous functions, \binenhance{} successfully reduces false positives in HermesSim.

Figure~\ref{fig:real} shows the distribution of AP scores for each method in 12 CVE vulnerabilities. The internal code semantic methods, such as Gemini and Asteria, fail to recall many vulnerabilities and low AP scores. It means that even if they recall vulnerabilities, the ranking of these vulnerabilities is not high. In our final false analysis of \binenhance{}, two primary factors were identified. First, \binenhance{} is partly contingent on the performance of HermesSim. For example, the $curl\_mvaprintf$ function is recalled as the $curl\_maprintf$ function of ${CVE-2016-8618}$. The cosine similarity score calculated by HermesSim is more than its homologous function. Second, certain non-homologous functions share similar external environments, yet these instances are relatively rare. In a word, \binenhance{} can enhance the semantics contained in the embeddings generated by the original model in most cases by introducing external environment semantics (EESG in Section~\ref{3.2.1}).

\begin{tcolorbox}
\vspace{-5pt}
\textbf{Answer to RQ.6.} \binenhance{} identified 101 1-day vulnerabilities in 37 firmware images with 67.9\% MAP scores, which is 12 more vulnerabilities detected than HermesSim, and a MAP score of 7.7\% higher.
\vspace{-5pt}
\end{tcolorbox}

\section{Related Work}

In the field of Binary Code Search, various techniques have been proposed to tackle the challenge of identifying similarities between the binary code. We categorize the related work into two main groups: methods only utilizing internal code features, and those combining internal features and call graphs (CG).

\textbf{Internal Code Semantic Methods (ICSM).}
Several prior studies have concentrated on analyzing internal code features of binary code snippets to detect similarity. 
These methods largely utilize features like text semantics, control flow graphs (CFGs), or abstract syntax trees (ASTs). 
For instance, text semantic-based approaches such as asm2vec \cite{ding2019asm2vec}, InnerEYE \cite{zuo2018neural}, Trex \cite{pei2020trex}, jTrans \cite{wang2022jtrans}, and Inter-BIN \cite{song2022inter}, interpret programming languages through natural language processing techniques to compute similarities. 
Similarly, Gemini \cite{xu2017neural} and ISRD \cite{xu2021interpretation} use CFGs to analyze structural features in binary code for similarity assessment. 
Asteria \cite{yang2021asteria} and others \cite{wang2020detecting, buch2019learning} involve the use of ASTs for function matching, focusing on syntactic structure analysis.  
HermesSim~\cite{he2024code} and VulHawk\cite{luovulhawk} convert binary code into IR and construct semantic graphs (e.g., SOG) to encode functions.
Despite their effectiveness in some scenarios, these methods can not fully capture the complete semantics of the binary code, limiting their effectiveness in certain complex cases (inlining, etc.).

\textbf{The combination of CG and ICSM Methods.}
To overcome the limitations inherent in solely internal code semantic methods, recent studies have explored incorporating external semantic features, especially function call graphs. 
These integrated approaches aim to improve binary code similarity detection. 
$\alpha$diff \cite{liu2018alphadiff} represents an innovative example, using a two-dimensional vector for each node's in/out-degree on the function call graph to represent function features. 
Furthermore, BMM \cite{guo2022exploring} is another noteworthy work that extends the idea of integrating call graphs with internal semantic information. 
This approach constructs a representation of binary code by combining a control flow graph, data flow graph, and function call graph. 
By capturing both the structural and behavioral aspects of the code, BMM \cite{guo2022exploring} demonstrates superior performance in identifying code similarities across diverse binary.

In this paper, we propose a novel method that goes beyond the existing approaches by integrating both internal semantic information and external semantic information for binary code search. By combining the comprehensive understanding of the code's internal structure with the contextual knowledge derived from external semantic information, our method aims to capture a more holistic representation of binary code snippets. This integration allows us to leverage both the local semantics within a code snippet and the broader inter-function relationships, leading to improved accuracy and robustness in detecting similarities between binary codes.

\section{DISCUSSION}

In this section, we discuss the limitations and future research of the \binenhance{} framework. 
\binenhance{} significantly enhances existing binary code search methods by incorporating external semantics. 
It demonstrates stable performance improvements across various scenarios, including cross-architecture, cross-optimization options, function inlining, and function pool expansion. 
The framework's compatibility with multiple baseline methods indicates its potential to enhance semantic learning within any function's internal features.

Our experiments reveal that integrating Address-Adjacency ($AA$) semantics leads to notable improvements. 
Nonetheless, the efficacy of $AA$ edge is constrained to reused code searches, lacking the capability to enrich non-reused code semantically. 
It may also be vulnerable to layout obfuscation techniques and could inadvertently integrate irrelevant code snippet semantics.
Furthermore, our 1-day detection experiments indicate a rarity of non-homologous functions with similar external environments, posing a risk of false positives. 
To address these challenges, future research will focus on developing adaptable algorithms for $AA$ edge creation, aiming to infuse meaningful location semantics more accurately.

Currently, many deep learning-based binary code search methods, particularly those employing natural language processing, lack mechanisms to manage complex scenarios such as cross-architecture. 
These methods typically either input assembly languages from various instruction sets into large self-learning models (e.g., TREX and Ordermaters \cite{yu2020order}) or focus on learning representations under a single instruction set (e.g., jTrans~\cite{wang2022jtrans} and Palmtree~\cite{li2021palmtree}). 
We posit the urgency of adopting Transformer-based models to comprehend cross-architecture semantics in assembly code, akin to multilingual models \cite{chi2020infoxlm}. Our future work will explore this approach and external semantic models with better learning capabilities.

\section{CONCLUSION}
In this paper, we propose a binary code search enhancement framework \binenhance{}, which enhances internal code semantic models with valuable external environment semantic information, thereby reducing the false positive and false negative. 
We design an EESG to model a function's external environment among homologous functions, linking the target function with the auxiliary functions and the target function with strings via diverse semantic dependencies. 
Moreover, We introduce whitening transformation and SEM to merge external semantics with internal ones, forming a comprehensive function representation.
We utilize data feature similarity to assist in calibrating \binenhance{}. 
Our approach demonstrates substantial improvements over existing methods in terms of performance and search speed in six tasks, as evidenced by experiments. 
Additionally, we have evaluated \binenhance{} against HermesSim in detecting 1-day vulnerabilities within a firmware dataset and show its advantages.

\section*{Acknowledgment}

We want to thank our shepherd and reviewers for their
insightful comments which highly improve our paper. Thanks to Associate Professor Xiaodong Gu for his help during the rebuttal period. The authors are
supported in part by the National Key Research and Development Program of China (2022YFB3103904) and Youth
Innovation Promotion Association CAS.




\bibliographystyle{IEEEtranS}
\bibliography{IEEEabrv,ref}

\end{document}